\documentclass[12pt]{article}
\usepackage{cyr}
\usepackage{epsf}
\usepackage{pazh}

\usepackage{amsmath}
\usepackage{amssymb}
\usepackage{graphicx}
\usepackage{indentfirst}
\usepackage{lscape}
\usepackage{afterpage}

\tightenlines

\sloppypar

\begin{document}
\baselineskip 21pt

\title{Do Disk Galaxies with Abnormally Low Mass-to-Light Ratios Exist?}
% ��������� �� ������ �������������� �������� \\

\author{\bf \hspace{-1.3cm}\copyright\, 2011 �. \ \
A. S. Saburova\affilmark{1*}, D. V. Bizyaev \affilmark{1,2**}, A. V. Zasov\affilmark{1***}}

\affil{
{\it Sternberg Astronomical Institute of Moscow State University, Universitetskii pr. 13, Moscow, 119992 Russia}$^1$\\
{\it Apache Point Observatory, New Mexico, USA}$^2$}

\vspace{2mm}
%\received{~~~~~~~~}
%\received{\today}
%\revised{}

\sloppypar \vspace{2mm} \noindent
We performed the photometric B, V and R observations of nine disk
galaxies that were
suspected in having abnormally low total mass-to-light (M/L) ratios
for their observed
color indices. We use our surface photometry data to analyze the
possible reasons for
the anomalous M/L. We infer that in most cases this is a result of errors in
photometry or rotational velocity, however for some galaxies we cannot
exclude the
real peculiarities of the galactic stellar population. The comparison
of the photometric
and dynamical mass estimates in the disk shows that the low M/L values
for a given color
of disks are probably real for a few our galaxies: NGC 4826 (Sab), NGC
5347 (Sab),
and NGC 6814 (Sb). The small number of such galaxies suggests that the
stellar initial
mass function is indeed universal, and that only a small fraction of
galaxies may have
a non-typical low-mass star depleted initial mass function. Such
galaxies require more
careful studies for understanding their star formation history.\\



\vfill
\noindent\rule{8cm}{1pt}\\
{$^*$ E-mail:$<$saburovaann@gmail.com$>$}\\
{$^{**}$ E-mail: $<$dmbiz@apo.nmsu.edu$>$}\\
{$^{***}$ E-mail: $<$a.v.zasov@gmail.com$>$}
\clearpage

\section{INTRODUCTION}
The mass-to-light ($M/L$) ratio is a very important parameter that is
determined by galactic stellar population (the distribution of stars in age, mass, and, to a
lesser extent, stellar metallicity), dust (through affecting the galaxy's
luminosity) and the relative mass of its nonstellar components.
The latters include the dark matter, which is usually comparable in mass to the stellar component
within the galaxy's optical boundaries (see, e.g., Khoperskov et al. 2010; Zasov et al.
2011; Bizyaev and Mitronova 2009), as well as the gas whose mass fraction can be significant
in late-type galaxies. The stellar population models computed for certain accepted initial 
mass function (IMF) obviously predict $M/L_B$  and
$M/L_V$ to be of the order of 1-10 solar units in the visible bands, and  $M/L_R$ or $M/L_K \sim
0.5-2$ solar units in the near infrared, in dependence of the relative number of young stars
(and therefore the galaxy's color) and the adopted IMF. The presence of the dark
matter or the gas along with internal extinction can only increase the
overal galactic 
$M/L$. Therefore, those galaxies whose total $M/L$ estimated within the optical
boundaries looks lower than that provided by the stellar population modeling using the
universal IMF (even ignoring any dark halo) are of especial interest.
Such candidate galaxies were selected from the sample of objects with known rotational
velocities by Saburova et al. (2009).

Since the reliable luminosity, color, and rotational velocity estimates are available
not for all selected objects, the conclusion about the $M/L$ anomaly for each specific galaxy
should be verified. An unusually low $M/L$ for a given color index, if confirmed, can
point to the existence of stellar population with a non-standard IMF: such a galaxy must have a
low relative number of stars with a mass less than the solar one because these stars
determine the total mass of stellar population once have a little affect on
the photometric features. This would imply the existence of special conditions for the formation of
the the bulk of its stars. Note that the questions about the IMF universality and the possibility of
a non-typical IMF shape are being actively discussed in the literature (see, e.g., Gilmore
2001; Kroupa 2002; Hoversten and Glazebrook 2008; Meureret al. 2009; Bastian et al. 2010;
Dabringhausen et al. 2010).

In this paper we present the results of our photometry for nine galaxies
selected from Saburova et al (2009) that were previously identified as presumably
having low total $M/L$ ratios within the optical radius $R_{25}$ (the radius of the 25
$^m/\Box''$ isophote in $B$) and analyze some possible sources of errors in the observed 
quantities.
We also compare the maximum rotation velocity of the disk component
$v_{max~disk}$, expected for a ''normal'' stellar population with the
maximum observed  rotation velocity $v_{max}$.
As a reference, we employ the model values of the
mass-to-light ratios for the stellar population with Salpeter IMF following the evolutionary
models by Bell \& de Jong (2001). The condition $v_{max~disk}
> v_{ max}$ suggests that the stellar population model with a regular IMFs is
inapplicable and that the the real stellar disk is lighter than it is predicted by the
stellar population models even if non-stellar components
(dark matter and gas) are ignored. To verify the total luminosity estimate as
well as the photometric parameters: color and inclination of the disk, we performed
photometric observations of nine galaxies in which we suspect the abnormally low
$M/L$.

\section{OBSERVATIONS AND DATA REDUCTION}
Photometric observations of the nine galaxies selected by Saburova et al. (2009) as
galaxies with presumably low $M/L$  were conducted at the 0.5-m
Apache Point Observatory telescope in the B,V, and R bands. We observed the
galaxies during photometric moonless nights in June, July, and October of 2009. 
Table \ref{table1} shows the observing
log.\footnote{The listed distances correspond to the Hubble constant $H =
75 km s^{-1} Mpc^{-1}$. For the nearby galaxy NGC 1569, we used the distance from Stil
\& Israel (2002).} The observing data were reduced in a standard way using MIDAS software
package and were corrected for the bias and flat field. The telescope's wide field of view
($40 '$) allowed the proper sky background subtraction using the starless areas
in the field. The photometric standard stars from Landolt
(1992, 2009) were observed during the same nights as the galaxies. The foreground
stars were removed from the galactic images and replaced by the mean
fluxes from the adjacent regions.
\begin{table}[h!]
 \caption{Log of observations. \label{table1}}
%\begin{center}
   \begin{tabular}{|c|c|c|c|c|c|}
  \hline
Galaxy& Date of observation & \multicolumn{3}{c|}{Band and exposure time, s}&D, Mpc\\
 (1)&(2)&\multicolumn{3}{c|}{(3)}&(4)  \\
\hline
& &B&V&R&\\
 \hline
NGC1569&15.10.09&3X500&3X300&3X300&2.2\\
\hline
NGC4016&14.06.09&500&200&300&49.2\\
NGC4016&15.06.09&2X500&2X300&2X300&49.2\\
\hline
NGC4214&15.06.09&2X500&2X300&2X300&7\\
\hline
NGC4826&15.06.09&2X500&2X300&2X300&7\\
\hline
NGC5347&15.06.09&500&300&300&32.5\\
NGC5347&21.06.09&500&300&300&32.5\\
\hline
NGC5921&15.06.09&500&300&300&25.2\\
NGC5921&21.06.09&2X500&2X300&2X300&25.2\\
\hline
NGC6814&21.06.09&2X500&2X300&2X300&20.7\\
\hline
NGC7743&21.06.09&500&300&300&24.7\\
\hline
UGC03685&15.10.09&3X500&3X300&3X300&26.8\\
\hline

\end{tabular}
\end{table}

\section{RESULTS OF PHOTOMETRIC OBSERVATIONS}
Table \ref{table2} shows the derived total magnitudes and
$(B-V)$ colors uncorrected for extinction. The total
magnitudes were calculated from the fluxes in an
aperture close to $R_{25}$ size. For the comparison, the table shows the apparent $B$ magnitudes
and $(B-V)$ colors from Hyperleda database.\footnote{http://leda.univ-lyon1.fr/.}
As follows from Table \ref{table2}, our total magnitudes and those from
Hyperleda agree well. NGC 4016 for which the
NED database\footnote{http://www.ned.ipac.caltech.edu/.} provides apparently 
erroneous value of $m_B$
corresponding to unrealistically low $(B-V)=0.03$
is an exception. We see a good agreement 
between the magnitudes and colors for our different
nights (no more than $0.1^m$  and
$0.04^m$ for the magnitudes and colors, respectively).

\begin{table}[h!]
\small \caption{Total magnitudes and colors. \label{table2}}
  \begin{center}
\begin{tabular}{|c | c| c| c| c| c|}
    \hline
{\scriptsize Galaxy}&\multicolumn{2}{c|}{$(B-V)_{tot}$}&\multicolumn{2}{c|}{$m_B$}&\multicolumn{1}{c|}{$m_R$}\\
\hline
&{\scriptsize this paper}& Leda & {\scriptsize this paper } & Leda & \\
\hline
 NGC1569&0.83&0.83&11.8&11.8&10.4\\
\hline
NGC4016&0.32&&14.5&13.8&13.8\\
\hline
NGC4214&0.47&0.46&10.4&10.2&9.5\\
\hline
NGC4826&0.84&0.84&9.37&9.30&7.98\\
\hline
NGC5347&0.75&0.76&13.5&13.4&12.1\\
\hline
NGC5921&0.69&0.66&11.7&11.7&10.5\\
\hline
NGC6814&0.9&0.8&12.2&12.1&10.6\\
\hline
NGC7743&0.92&0.90&12.5&12.4&11\\
\hline
UGC03685&0.7&&12.8&13.1&11.6\\
 \hline
 \end{tabular}
 \end{center}
\end{table}

Along from the total magnitudes we obtained the radial profiles
of isophotal flattening ($b/a$) in the R-band images azimuthally averaged in elliptical
rings (see Fig. 1). The radial surface brightness profiles in all three
$BVR$-bands are plotted in Fig. 2. The brightness profiles were corrected for extinction in
the Galaxy using Schlegel et al. (1998), but they were not corrected for the disk
orientation and internal extinction\footnote{Note that extinction inside
the
galaxies only slightly affects position of points on $M/L$--color diagram
relative to the model evolutionary dependencies because it both redden the 
colors
and increases the $M/L$, so it remains close
to the expected value for a given color.} (the profiles of NGC 1569 were not corrected for
the extinction in the Galaxy due to the uncertainty in the correction, see below). The radial
brightness and color profiles are shown in Figs. 2 and 3. The ranges of errors in Figs. 2
and 3 were calculated from the errors in the fluxes as in Vader \& Chaboyer
(1994) using the formula:
\begin{equation}\label{2}
\delta I = \sqrt{N_{tot}+(\delta n_{sky}A)^2},
\end{equation}
where $N_{tot}$ is the flux from the galaxy in aperture $A$,
and $\delta n_{sky}$  is the standard deviation from the mean
sky background determined in small apertures near the galaxy.

To calculate the stellar disk luminosity, we decomposed the images of the galaxies
into components using the BUDDA code developed by de
Souza et al. (2004) (version 2.2 which allows for the bulge, disk, 
bar, and central source components). In our analysis we evaluate the following 
components: the exponential
disk $\mu_d(r)=\mu_0+1.086r/r_d$, a bulge, and a bar with brightness distributed according
to the Sersic law $\mu_b(r)=\mu_e+c_n((r/r_e)^{1/n}-1)$. The decomposition results for the
R-band profiles are shown in Figs. 4 and 5, respectively. The structural parameters
determined for different bands are given in Table. \ref{table3}. It
contains the following data:\\
(1) galaxy name;\\
(2) the radial scale length of disk in arcsec;\\
(3) the central surface brightness of the disk
in $^m/\Box''$;\\
(4) the effective radius of the bulge in arcsec;\\
(5) the effective surface brightness of the bulge in
$^m/\Box''$;\\
(6) the Sersic index for the bulge;\\
(7) the effective radius of the bar in arcsec;\\
(8) the effective surface brightness of the bar in
$^m/\Box''$;\\
(9) the Sersic index for the bar;\\
(10) the photometric band.\\
\begin{figure}
\includegraphics[width=6.7cm,keepaspectratio]{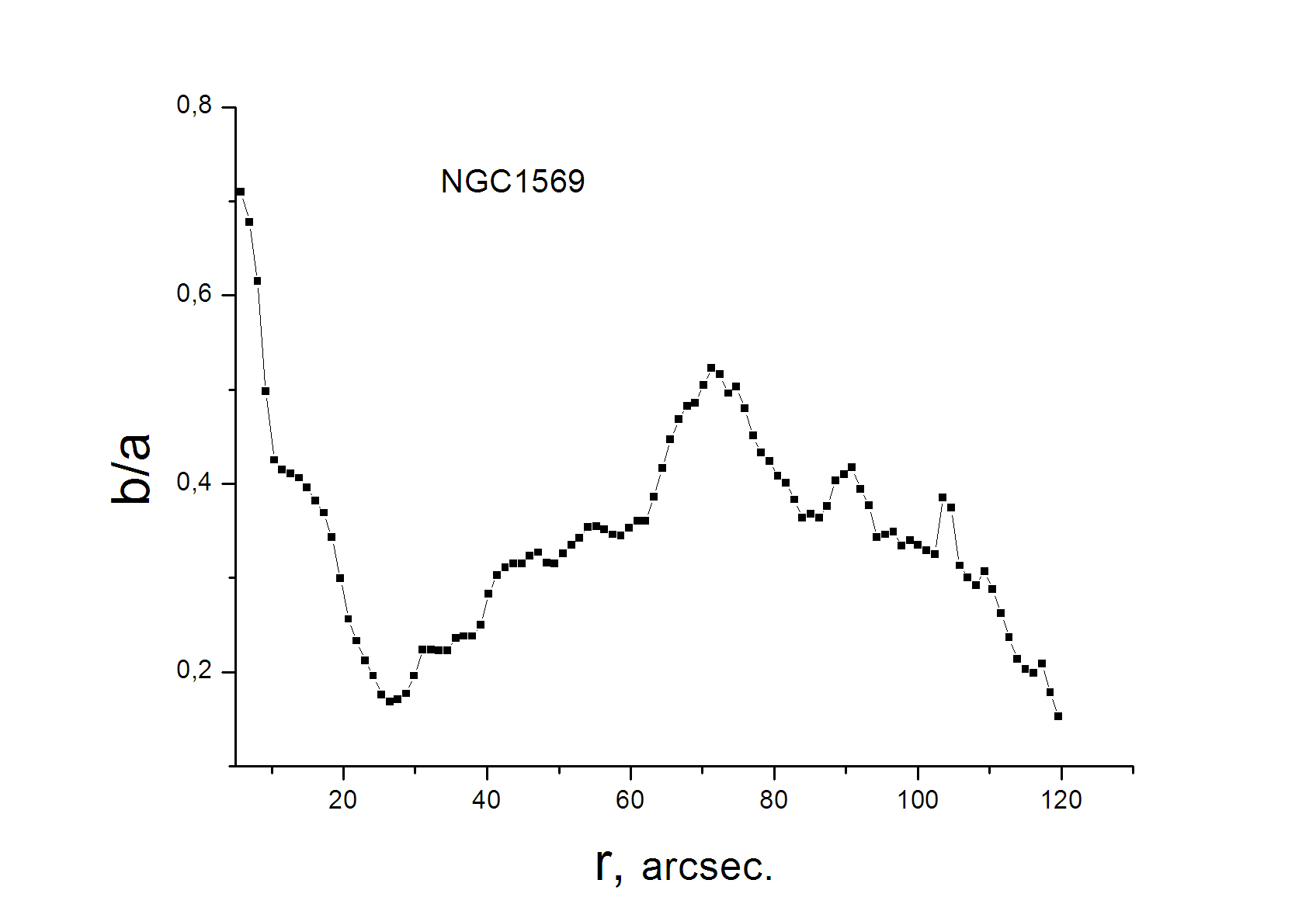}
\includegraphics[width=6.7cm,keepaspectratio]{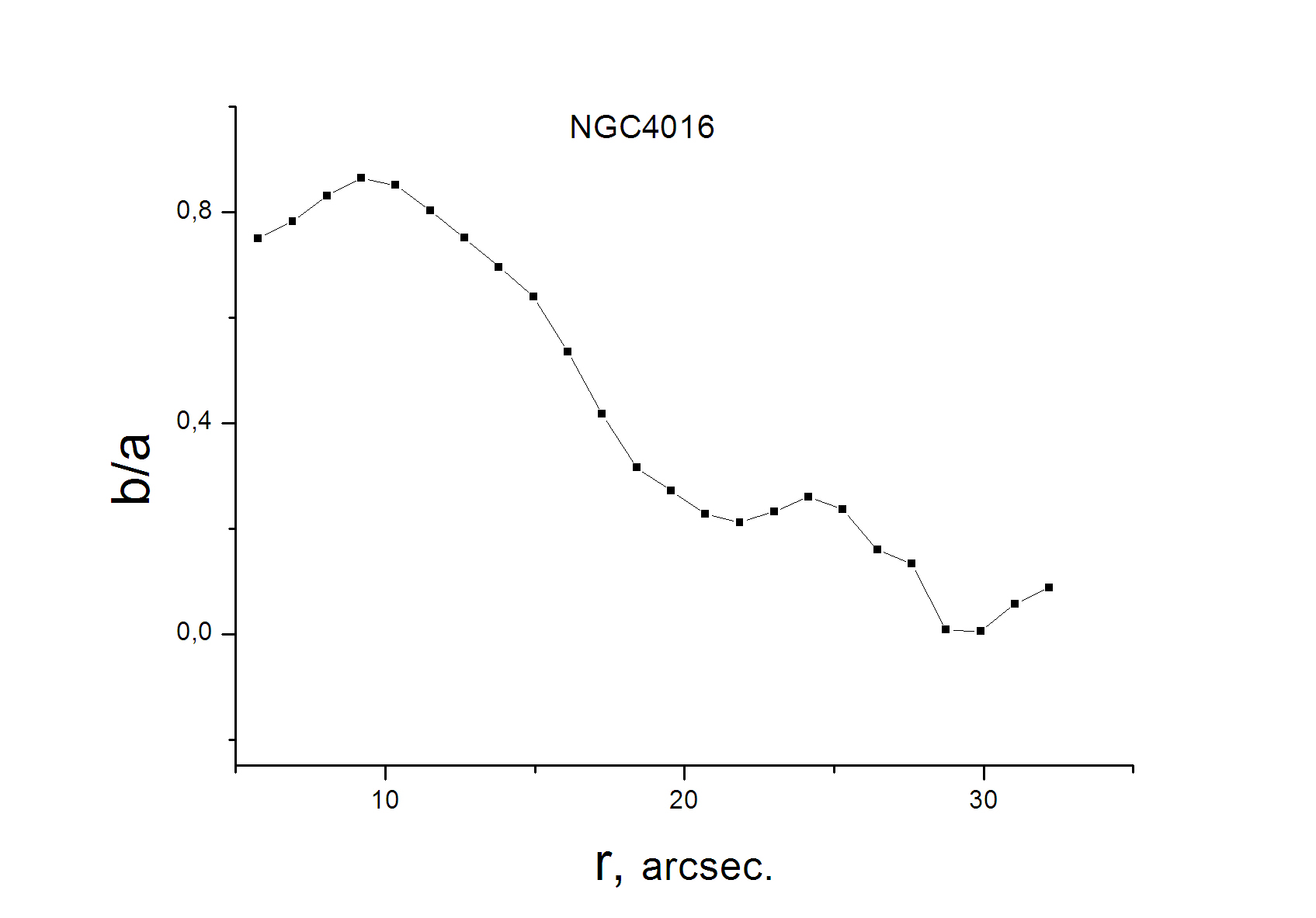}
\includegraphics[width=6.7cm,keepaspectratio]{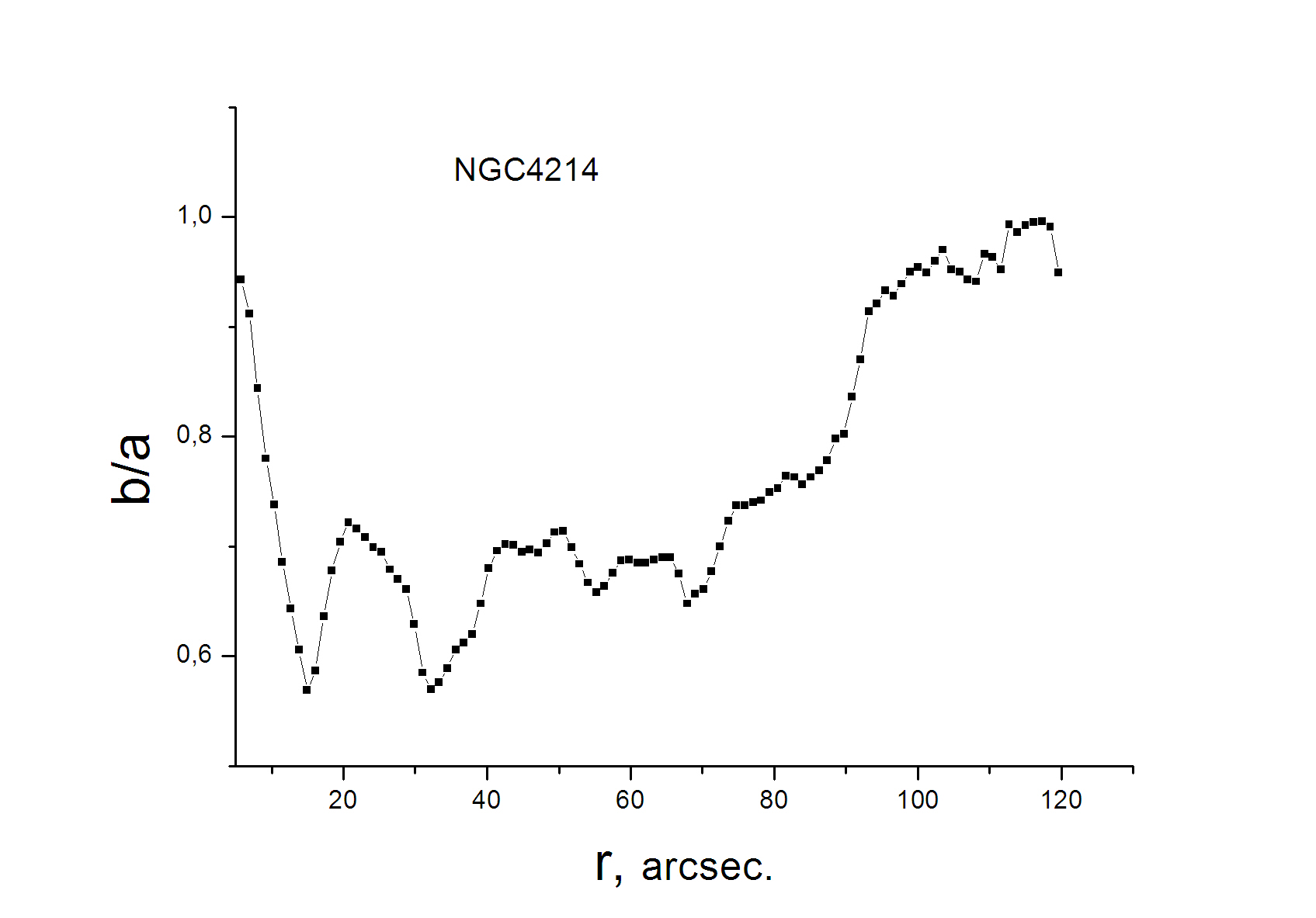}
\includegraphics[width=6.7cm,keepaspectratio]{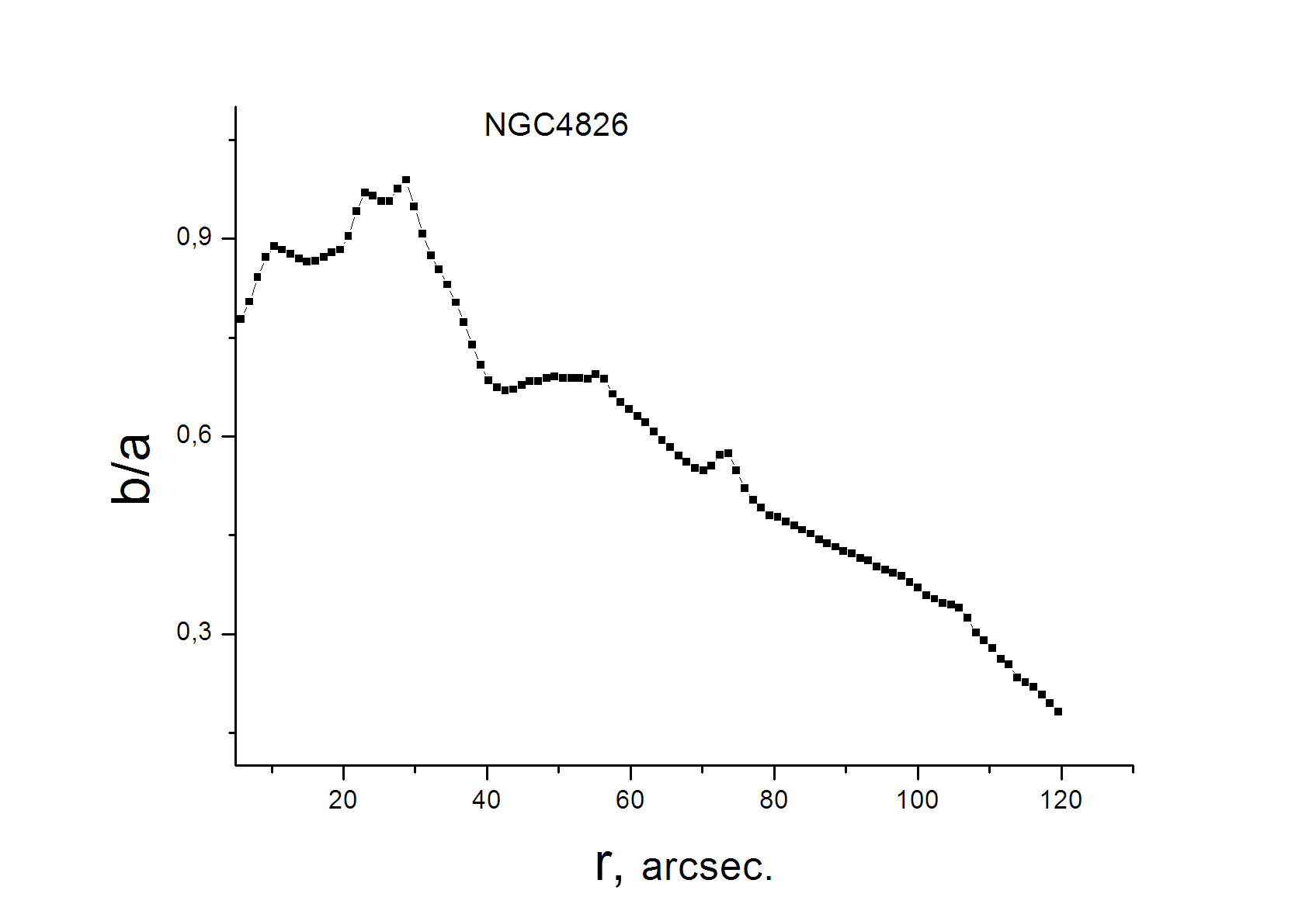}
\includegraphics[width=6.7cm,keepaspectratio]{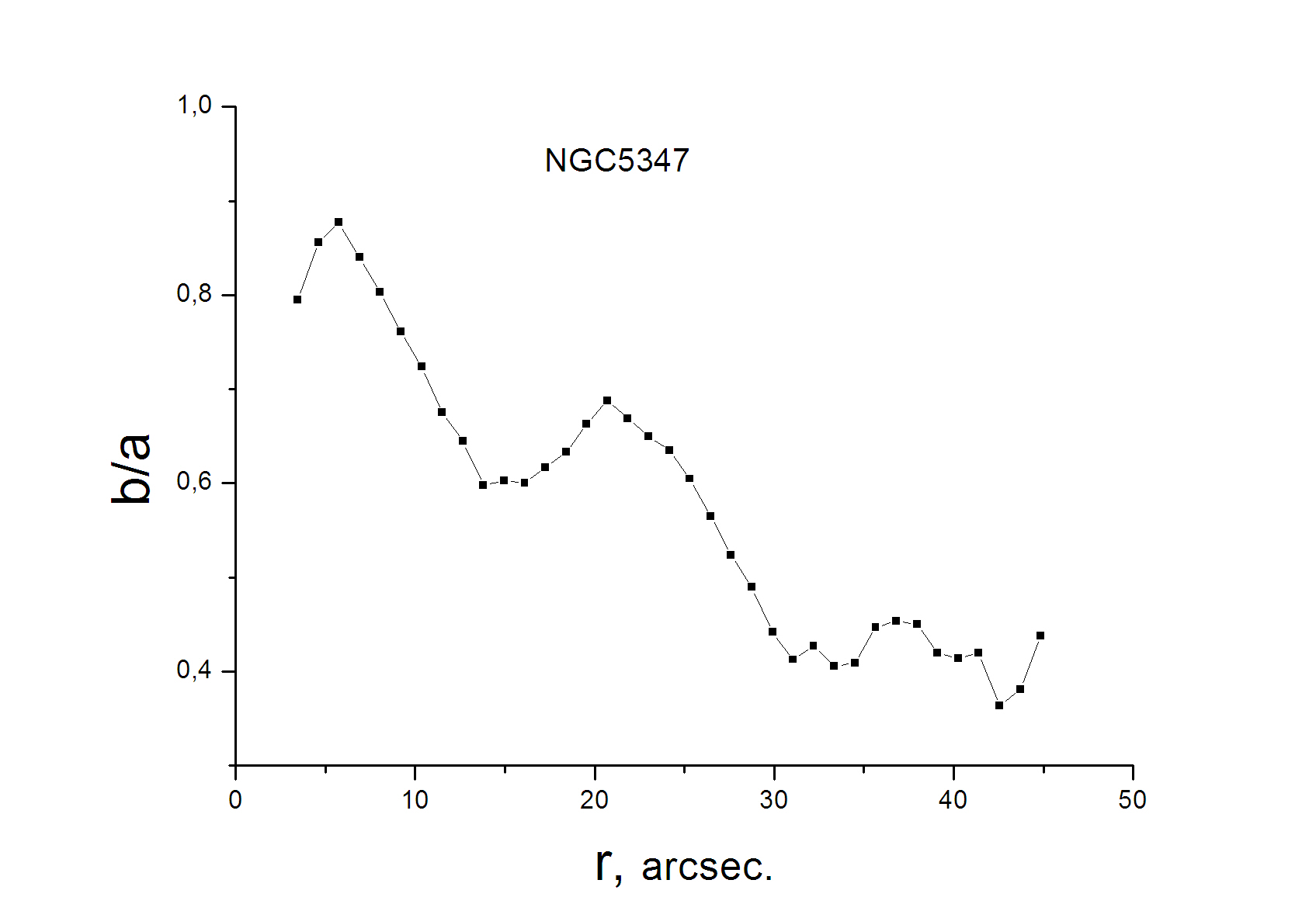}
\includegraphics[width=6.7cm,keepaspectratio]{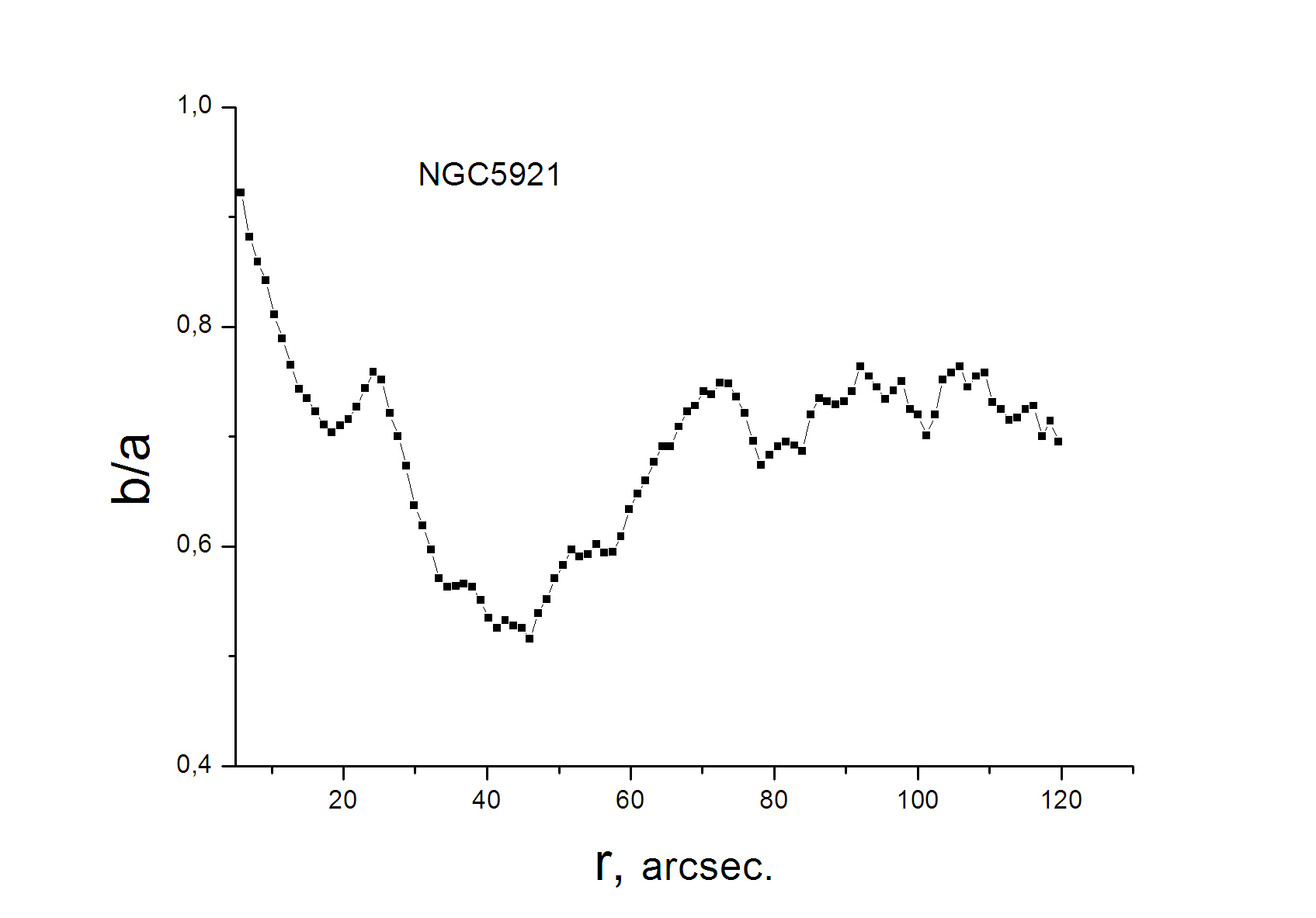}
\includegraphics[width=6.7cm,keepaspectratio]{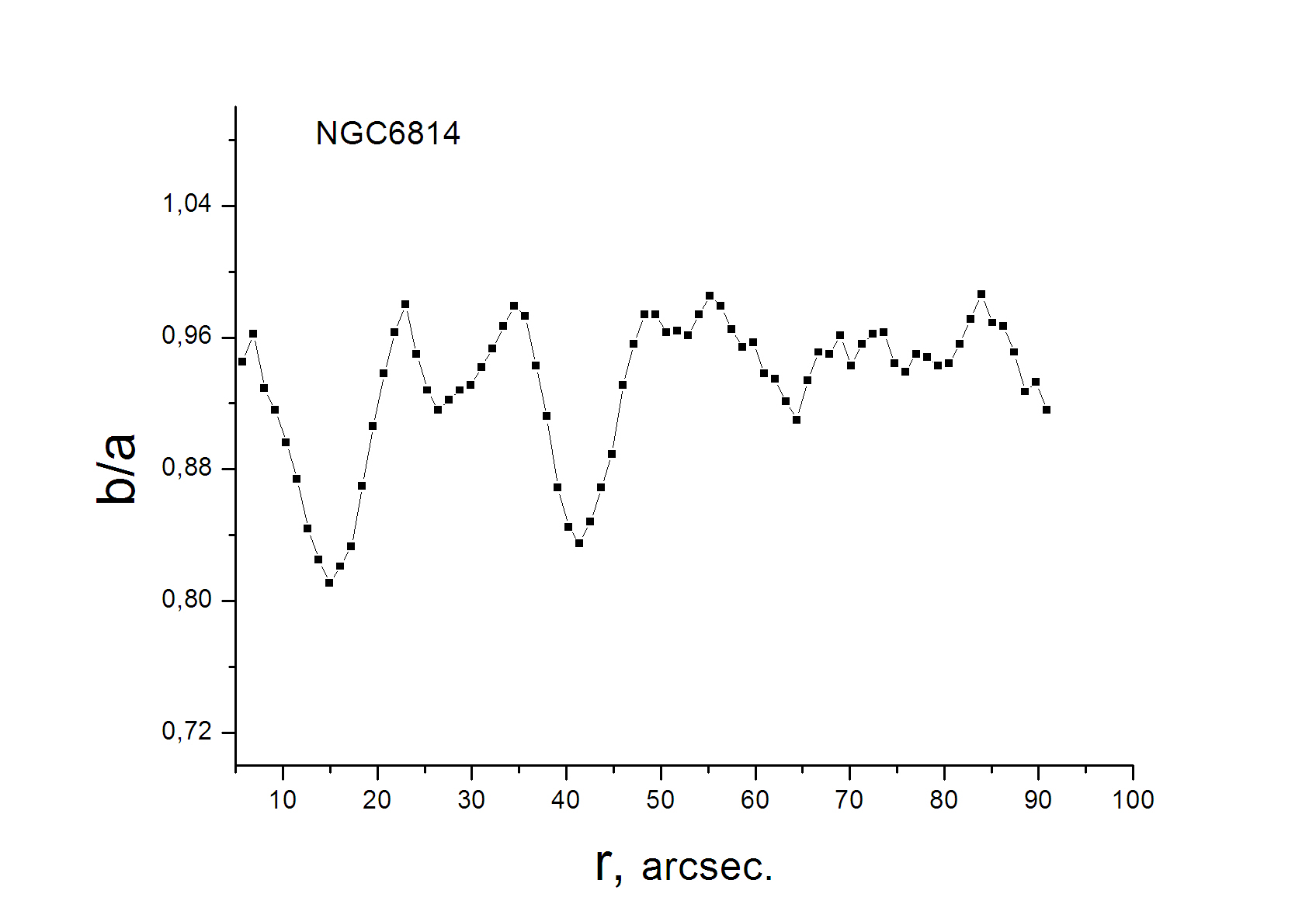}
\includegraphics[width=6.7cm,keepaspectratio]{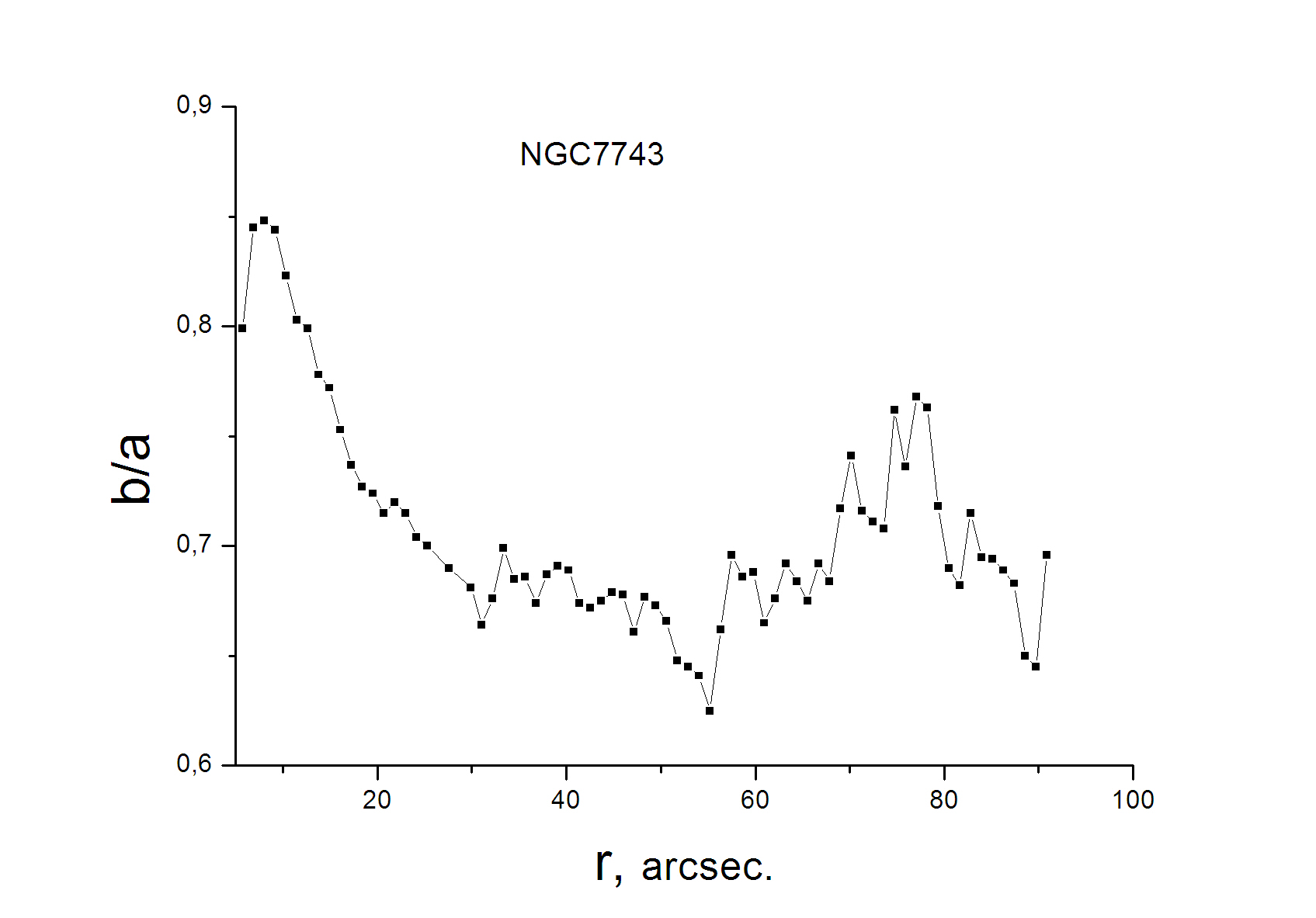}
\includegraphics[width=6.7cm,keepaspectratio]{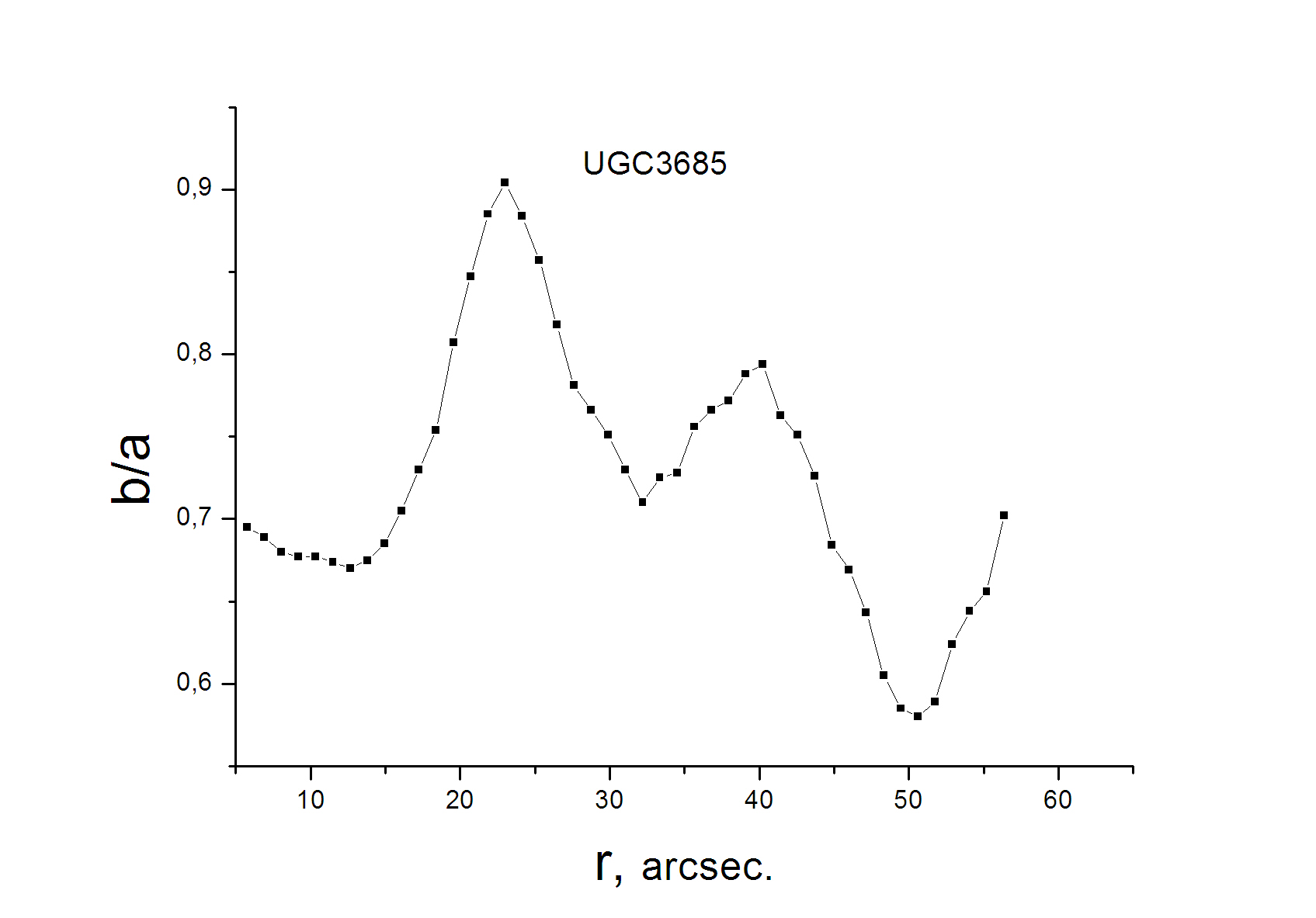}
\caption{Variations in the axial ratio of isophotes along the radius from R-band images.}
\end{figure}

\begin{figure}
\includegraphics[width=6.7cm,keepaspectratio]{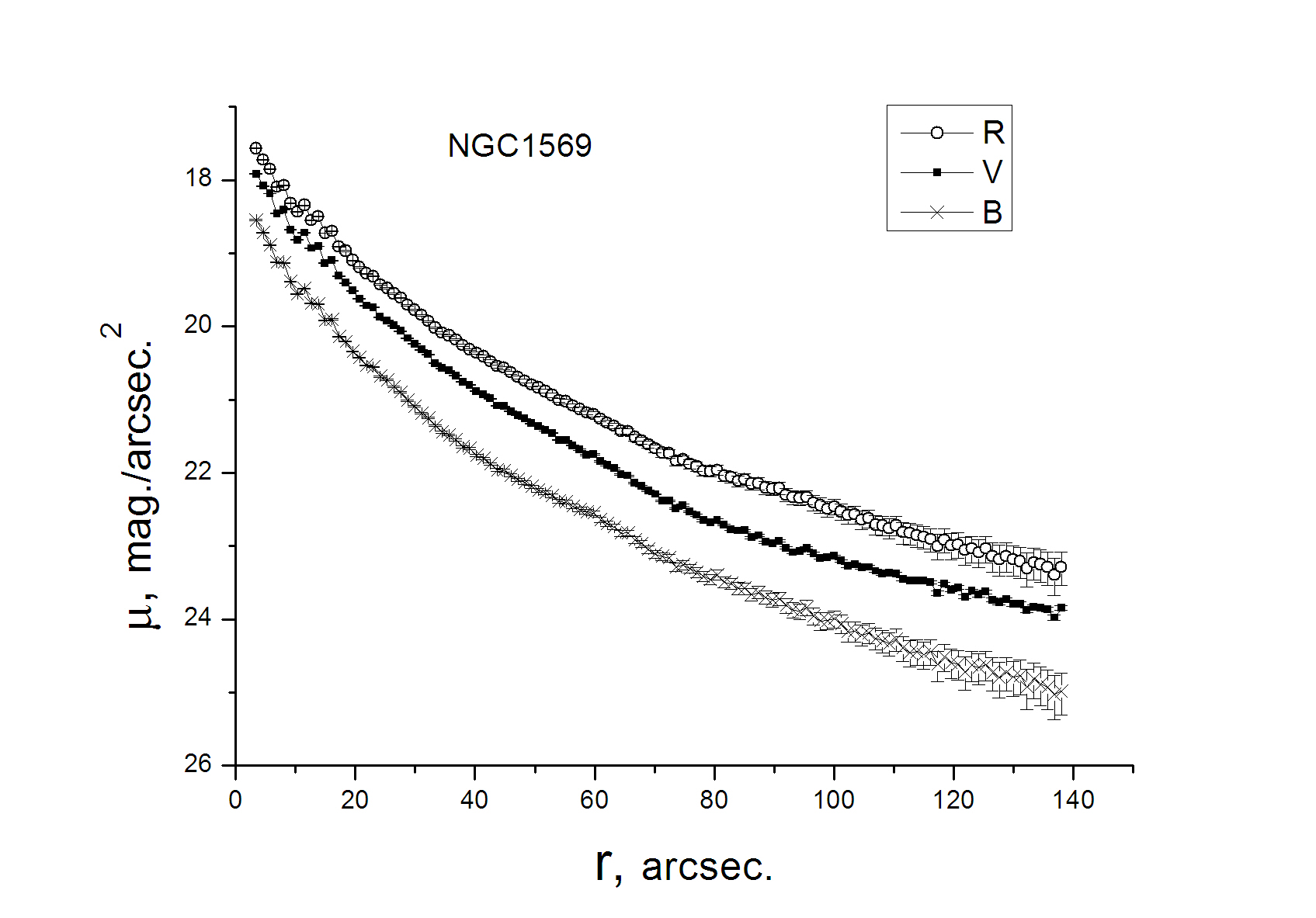}
\includegraphics[width=6.7cm,keepaspectratio]{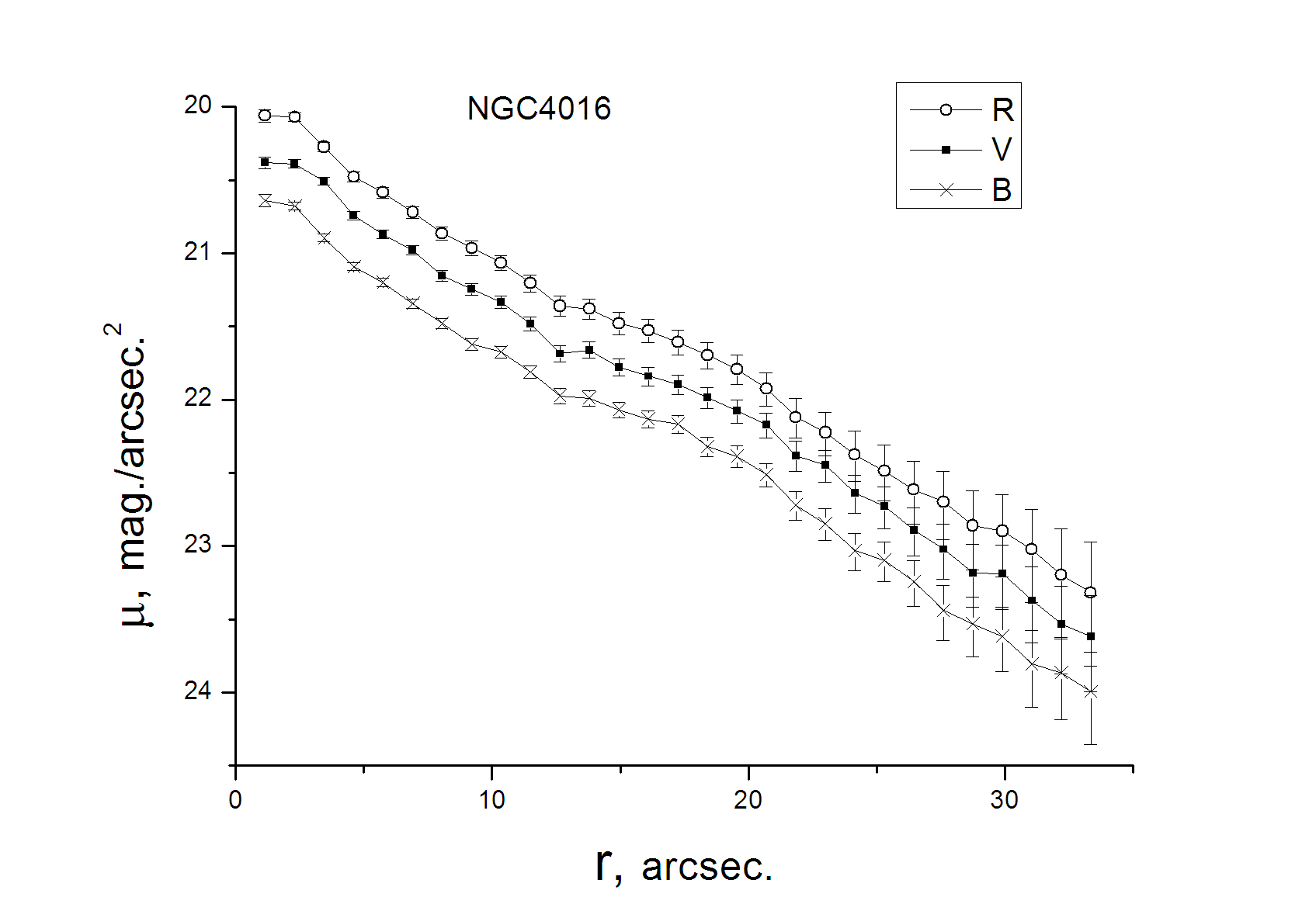}
\includegraphics[width=6.7cm,keepaspectratio]{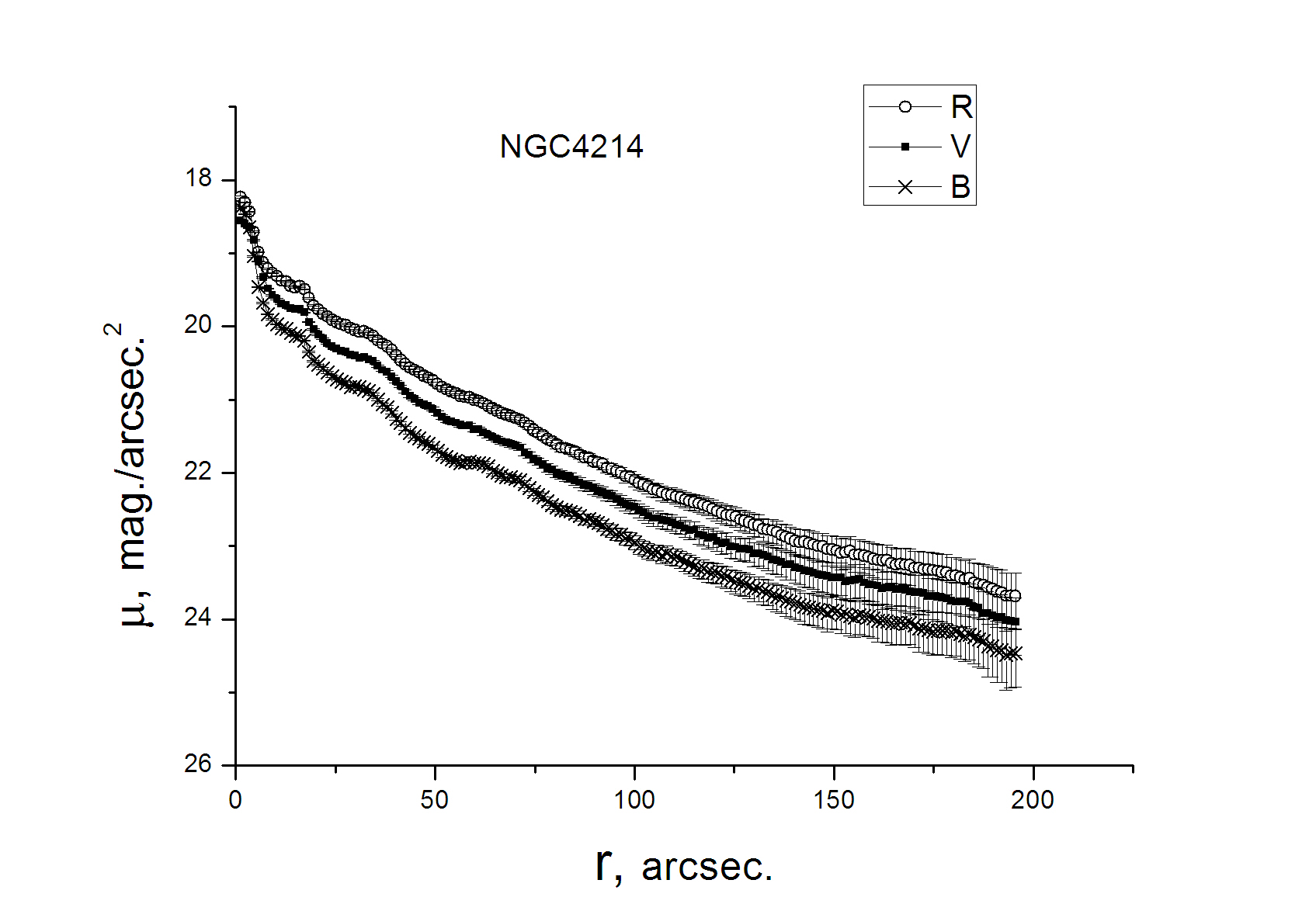}
\includegraphics[width=6.7cm,keepaspectratio]{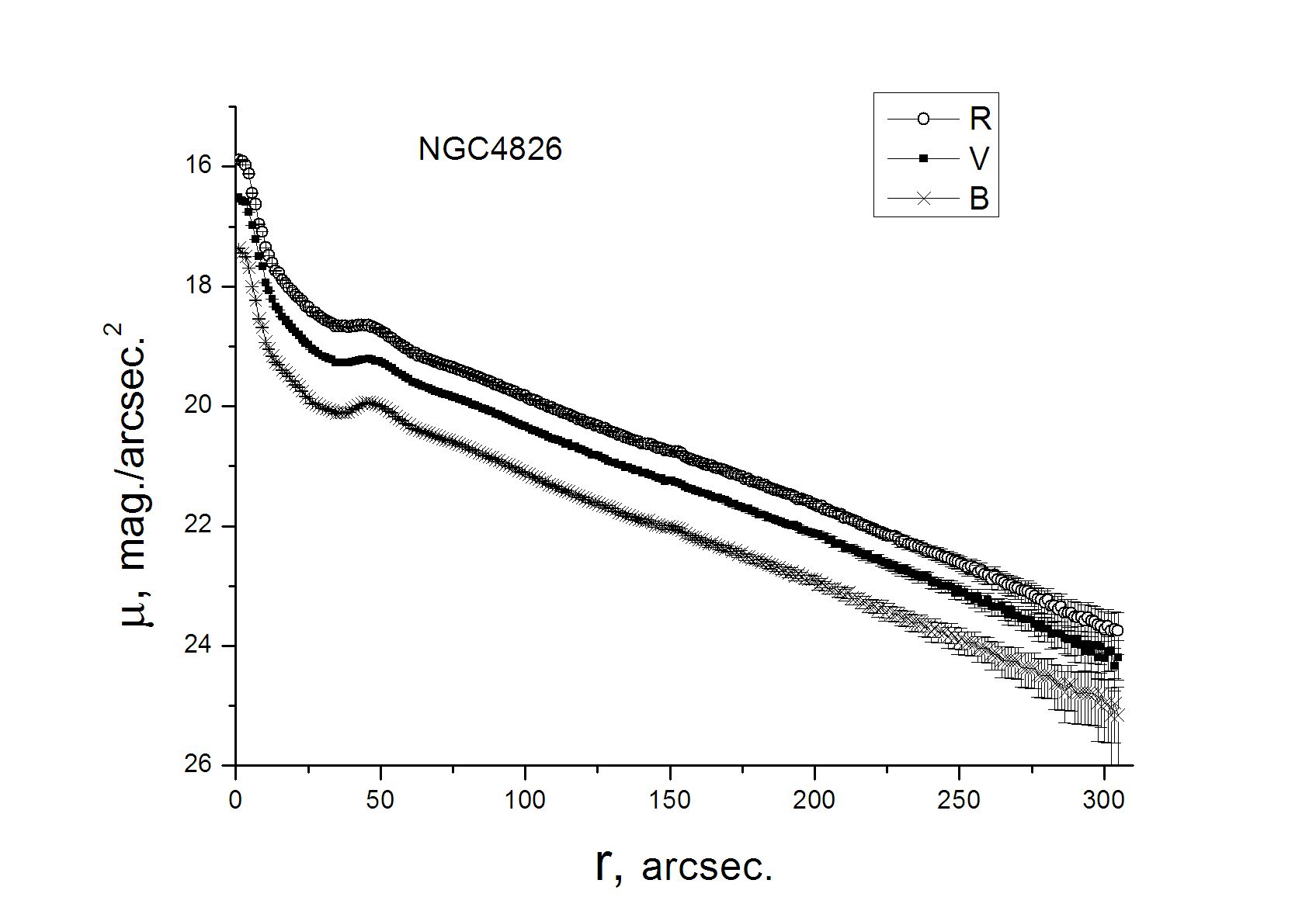}
\includegraphics[width=6.7cm,keepaspectratio]{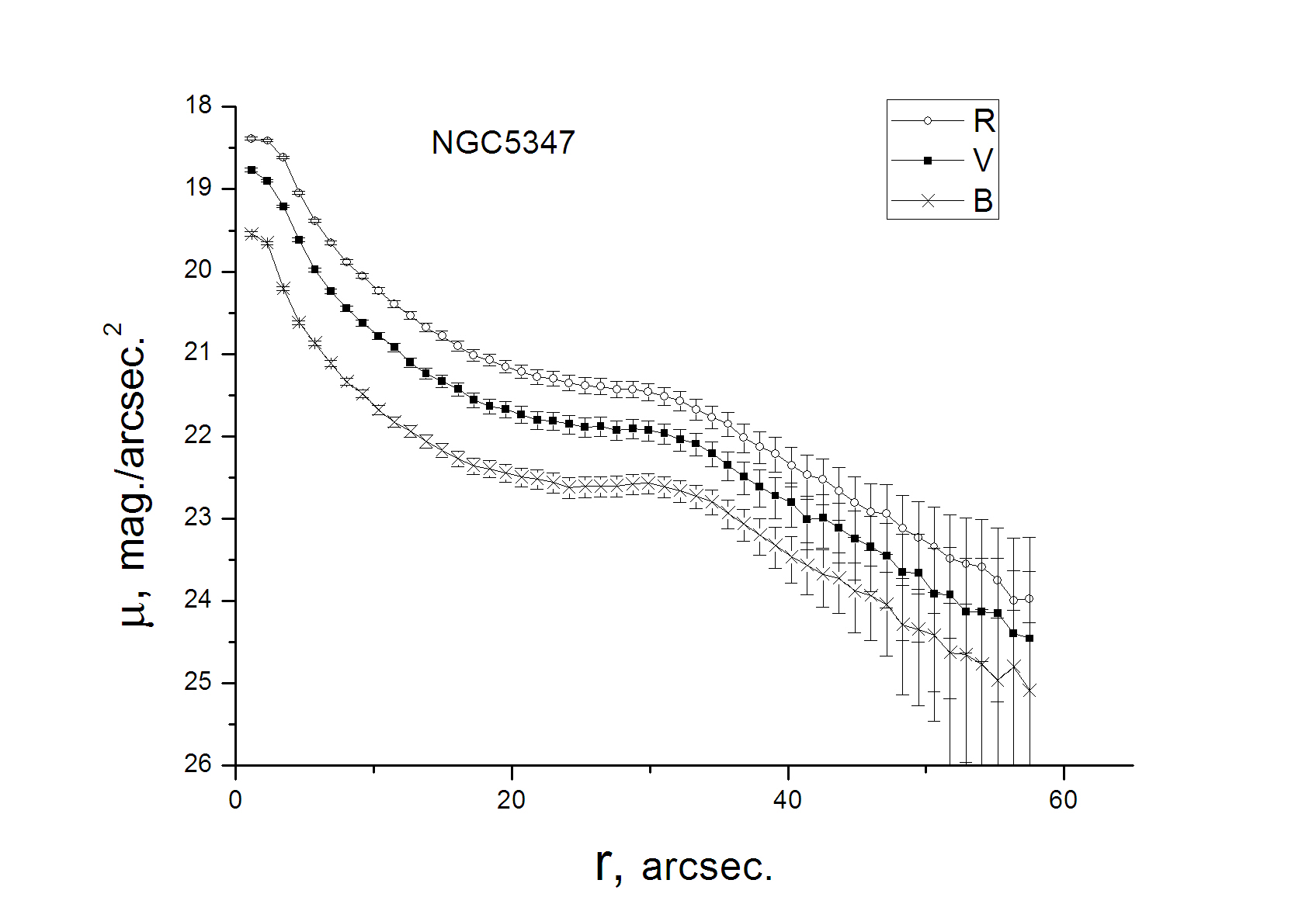}
\includegraphics[width=6.7cm,keepaspectratio]{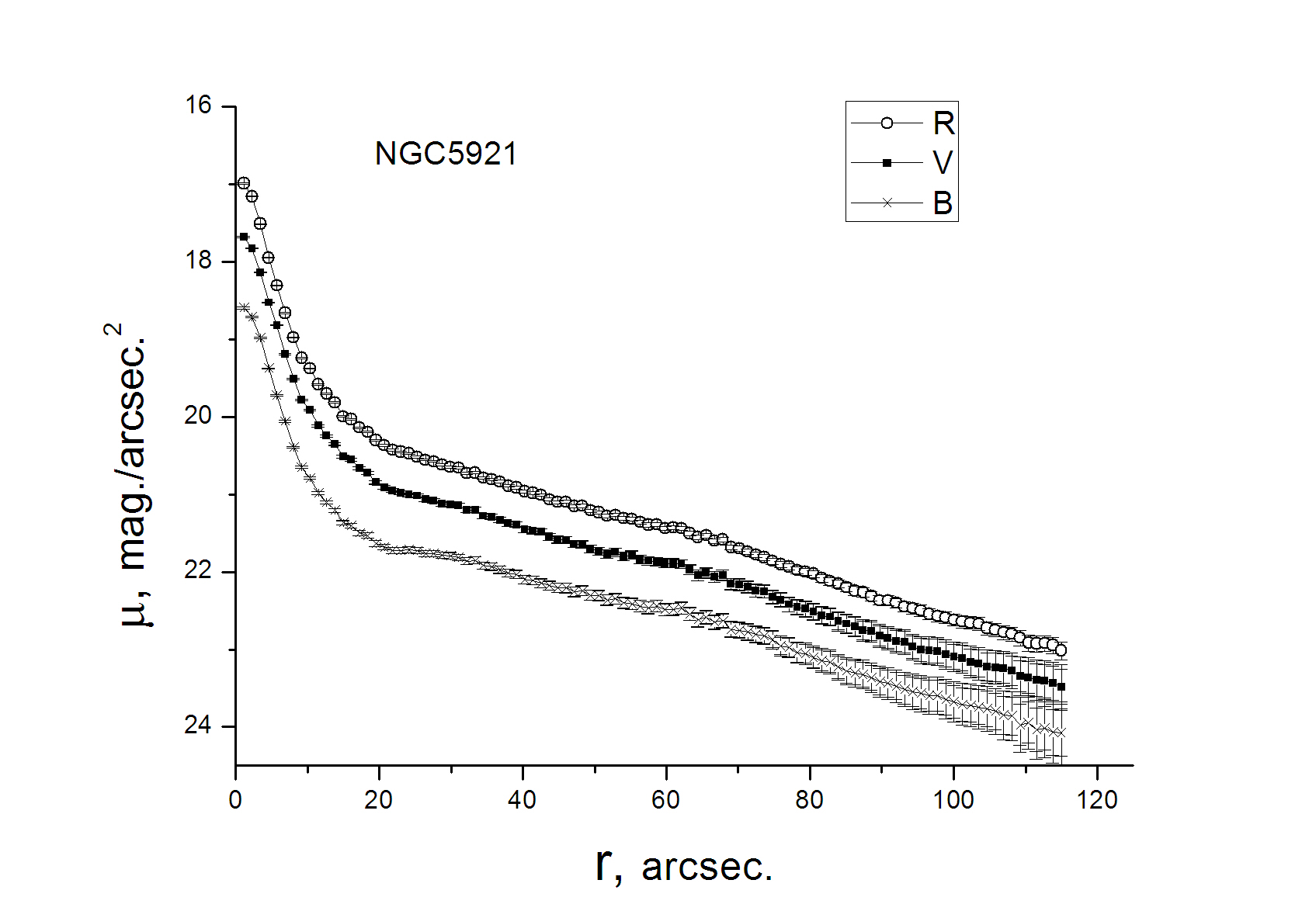}
\includegraphics[width=6.7cm,keepaspectratio]{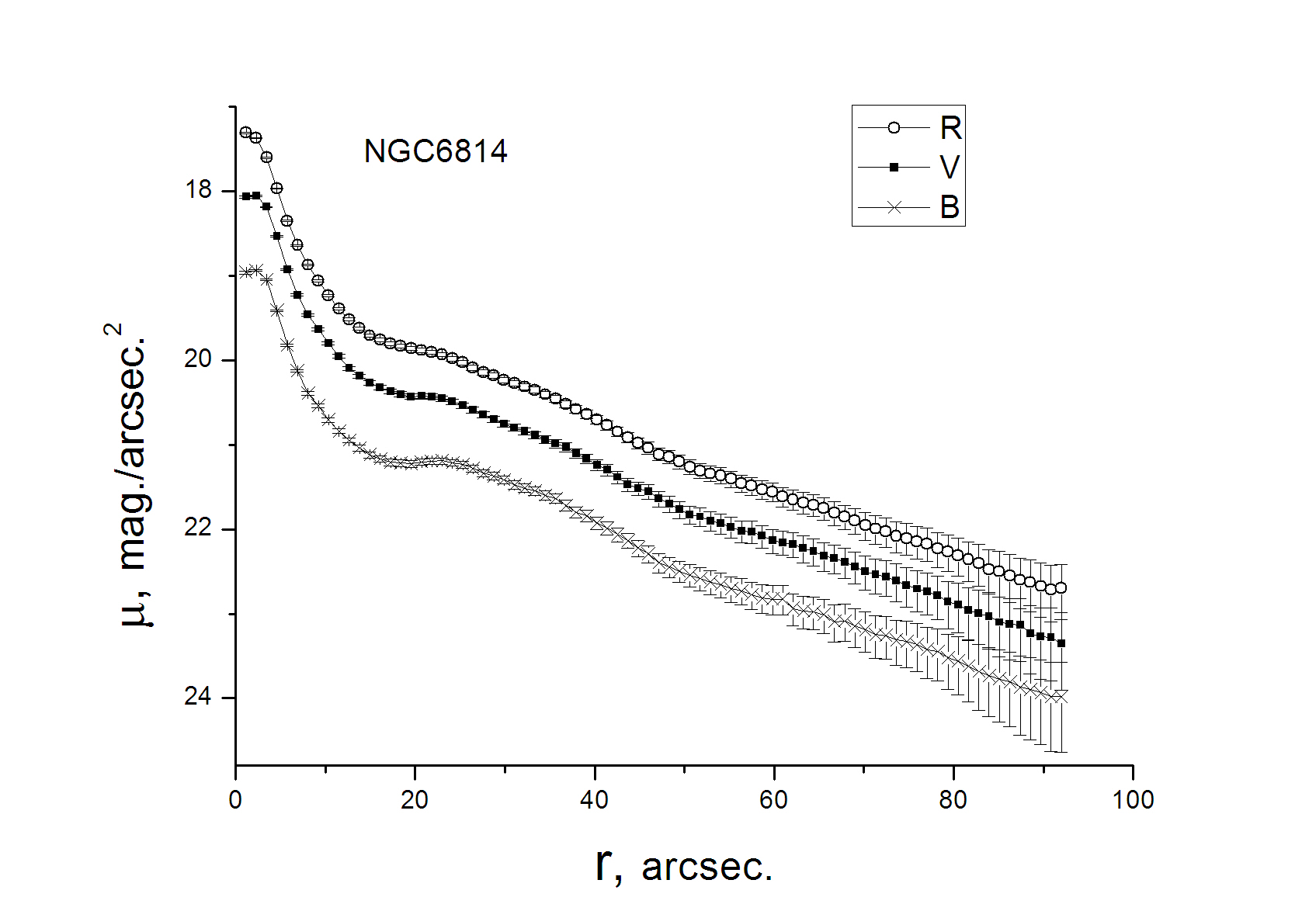}
\includegraphics[width=6.7cm,keepaspectratio]{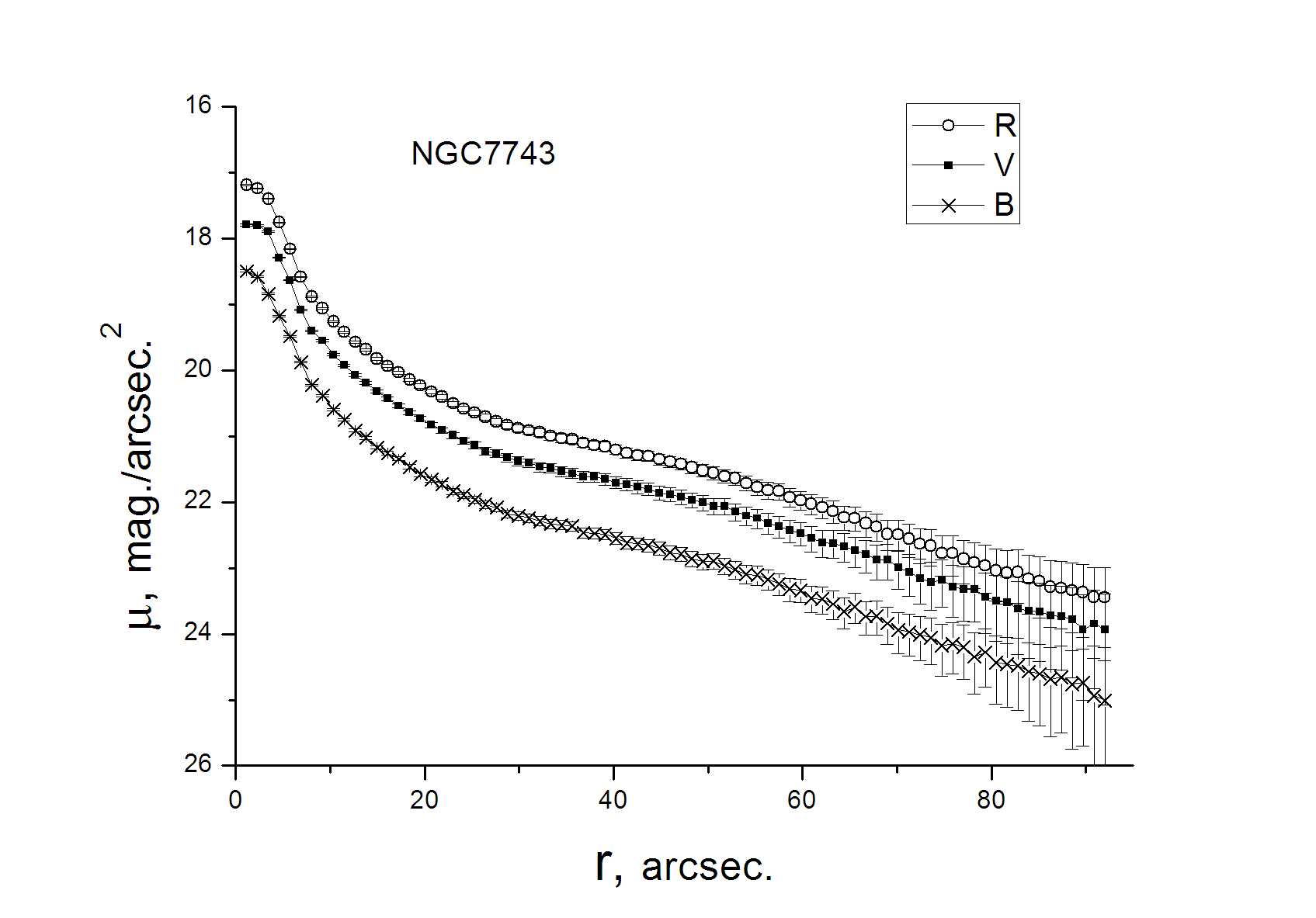}
\includegraphics[width=6.7cm,keepaspectratio]{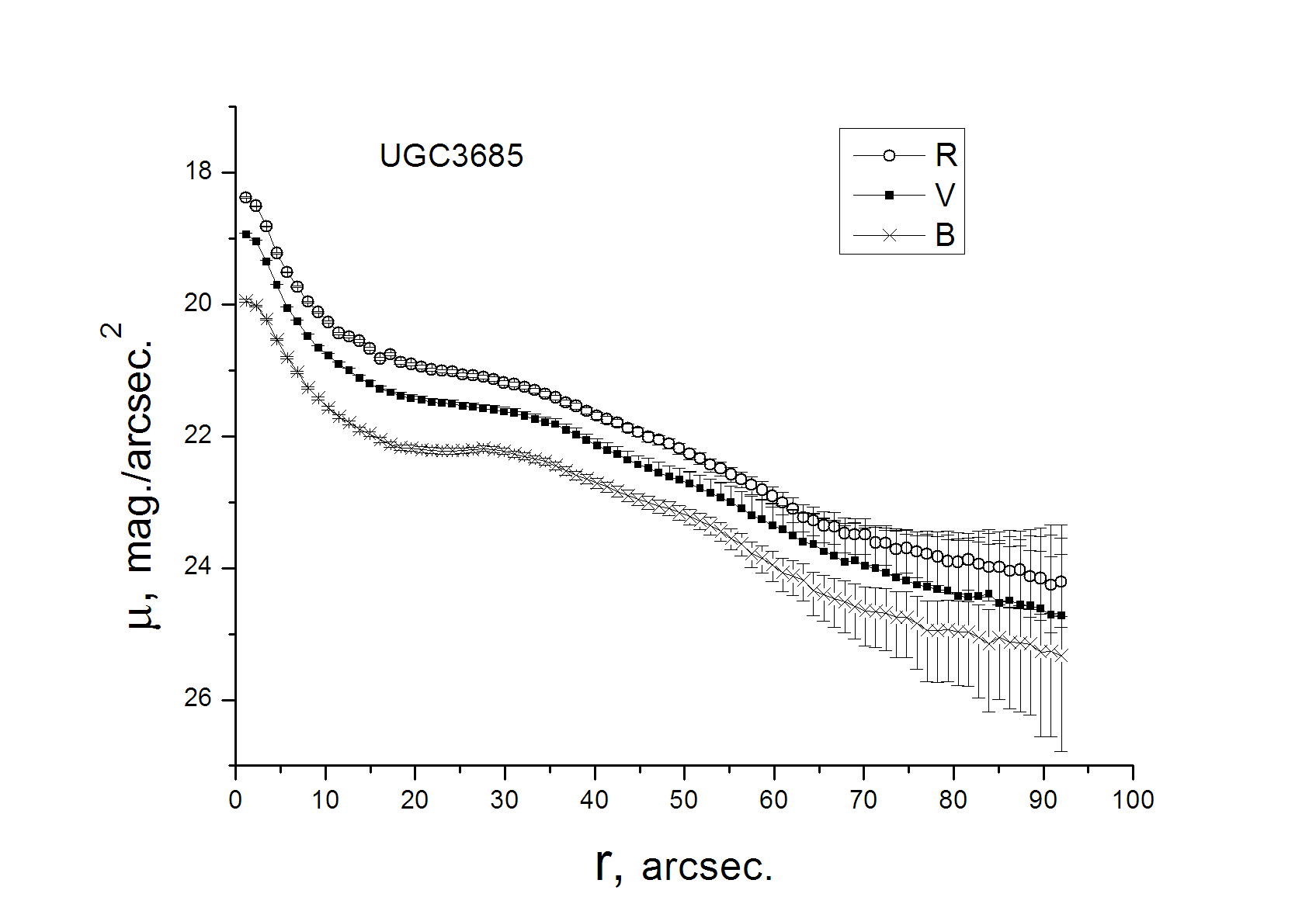}
\caption{Azimuthally averaged surface brightness profiles.}
\end{figure}

\begin{figure}
\includegraphics[width=6.7cm,keepaspectratio]{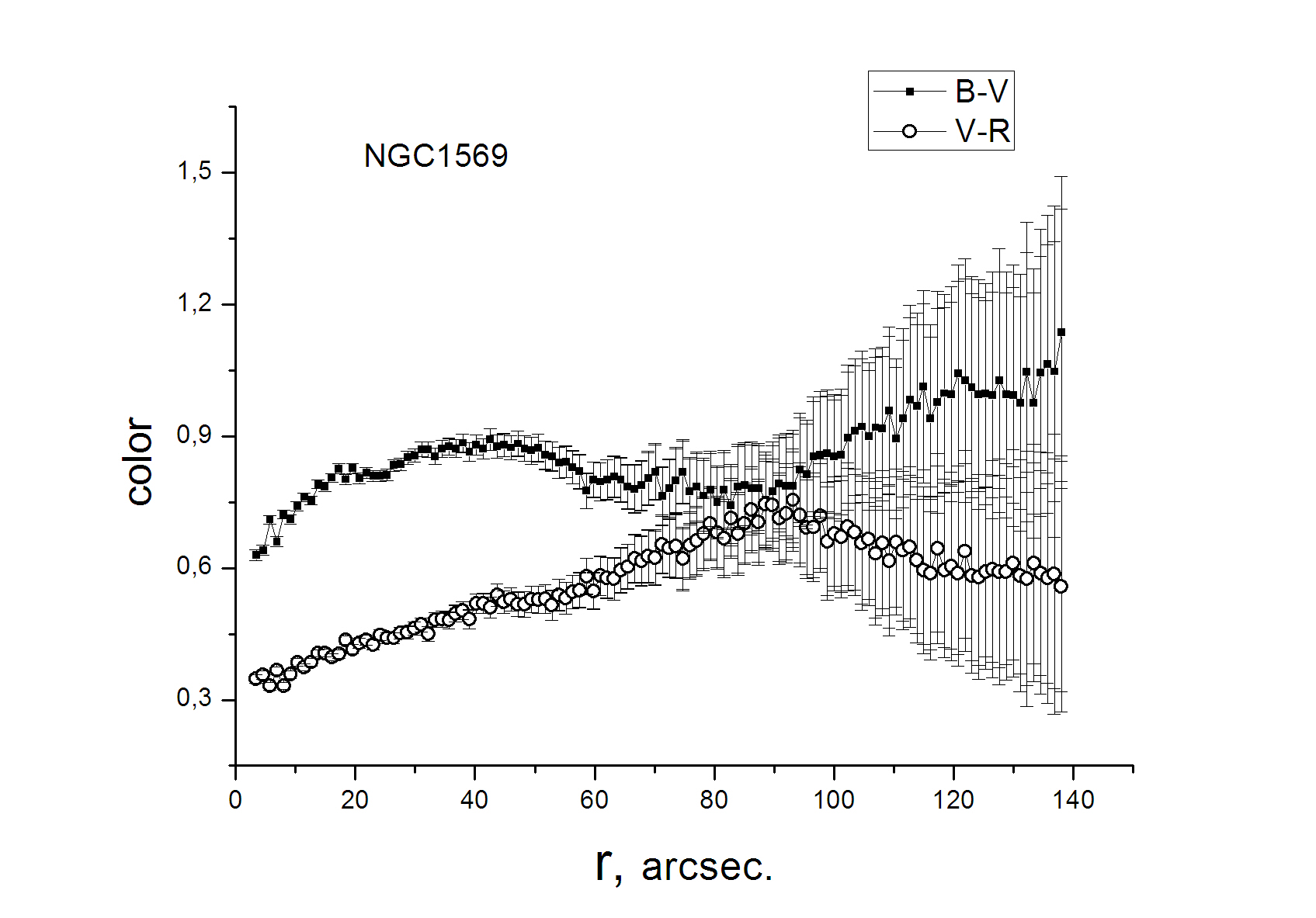}
\includegraphics[width=6.7cm,keepaspectratio]{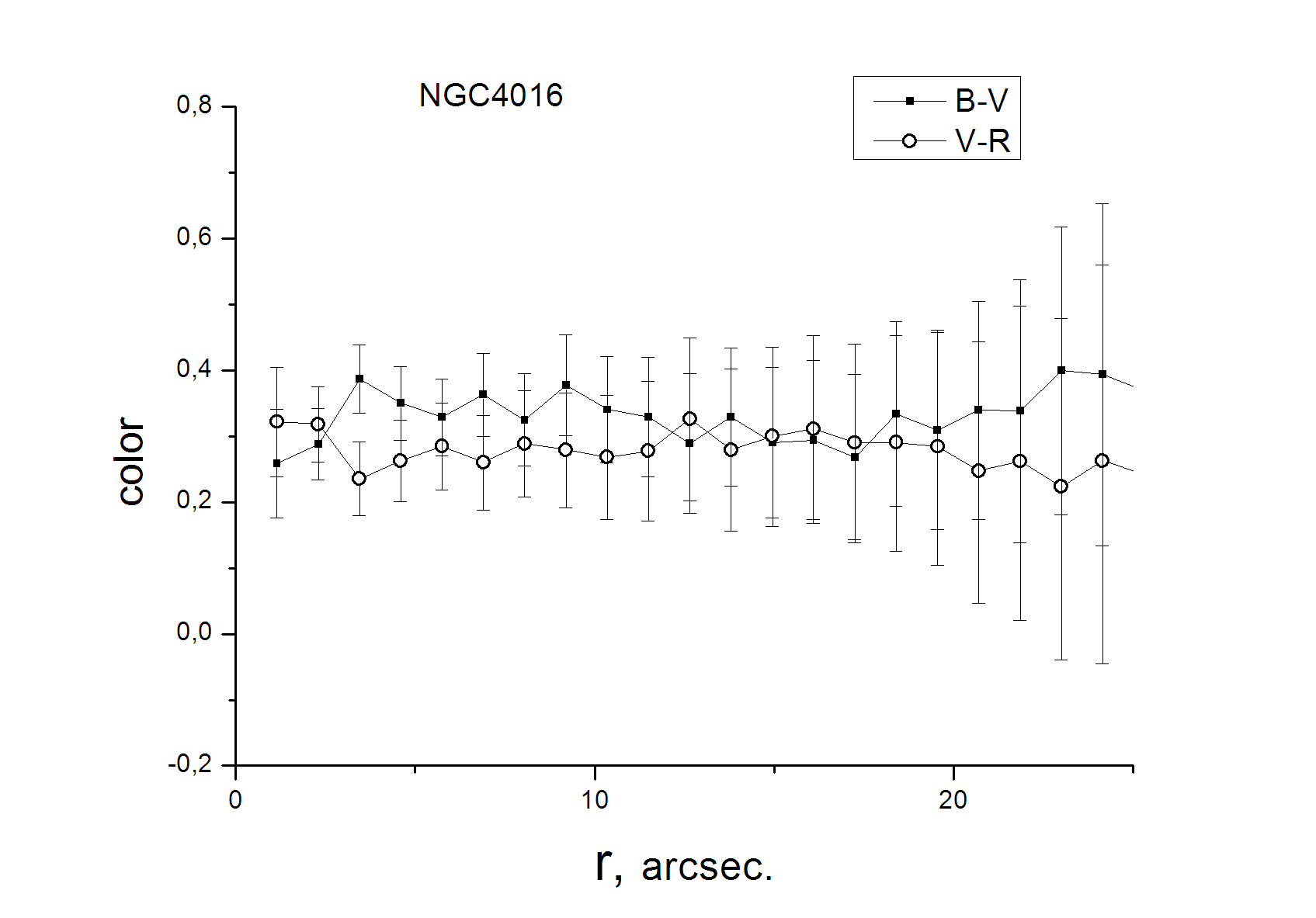}
\includegraphics[width=6.7cm,keepaspectratio]{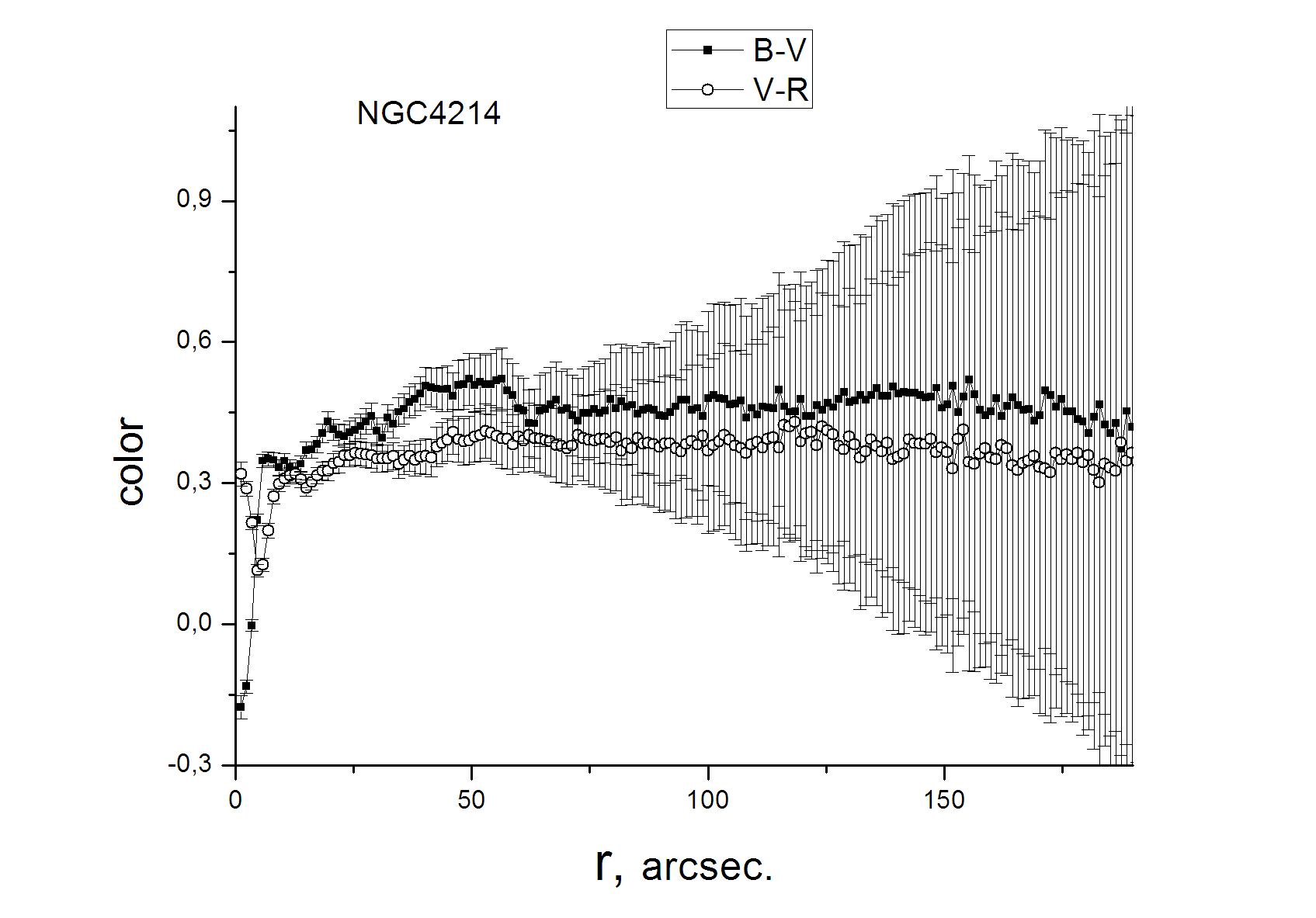}
\includegraphics[width=6.7cm,keepaspectratio]{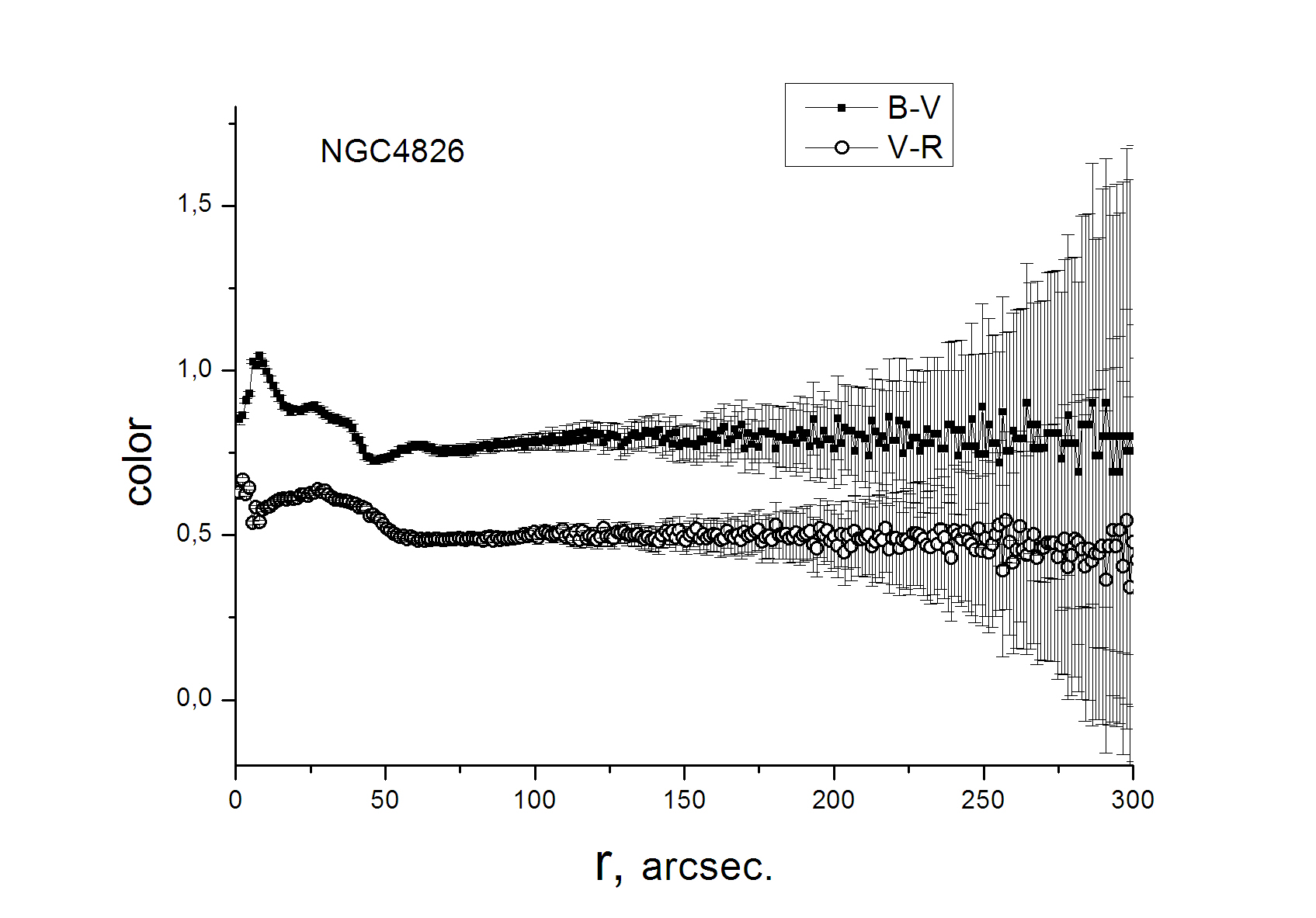}
\includegraphics[width=6.7cm,keepaspectratio]{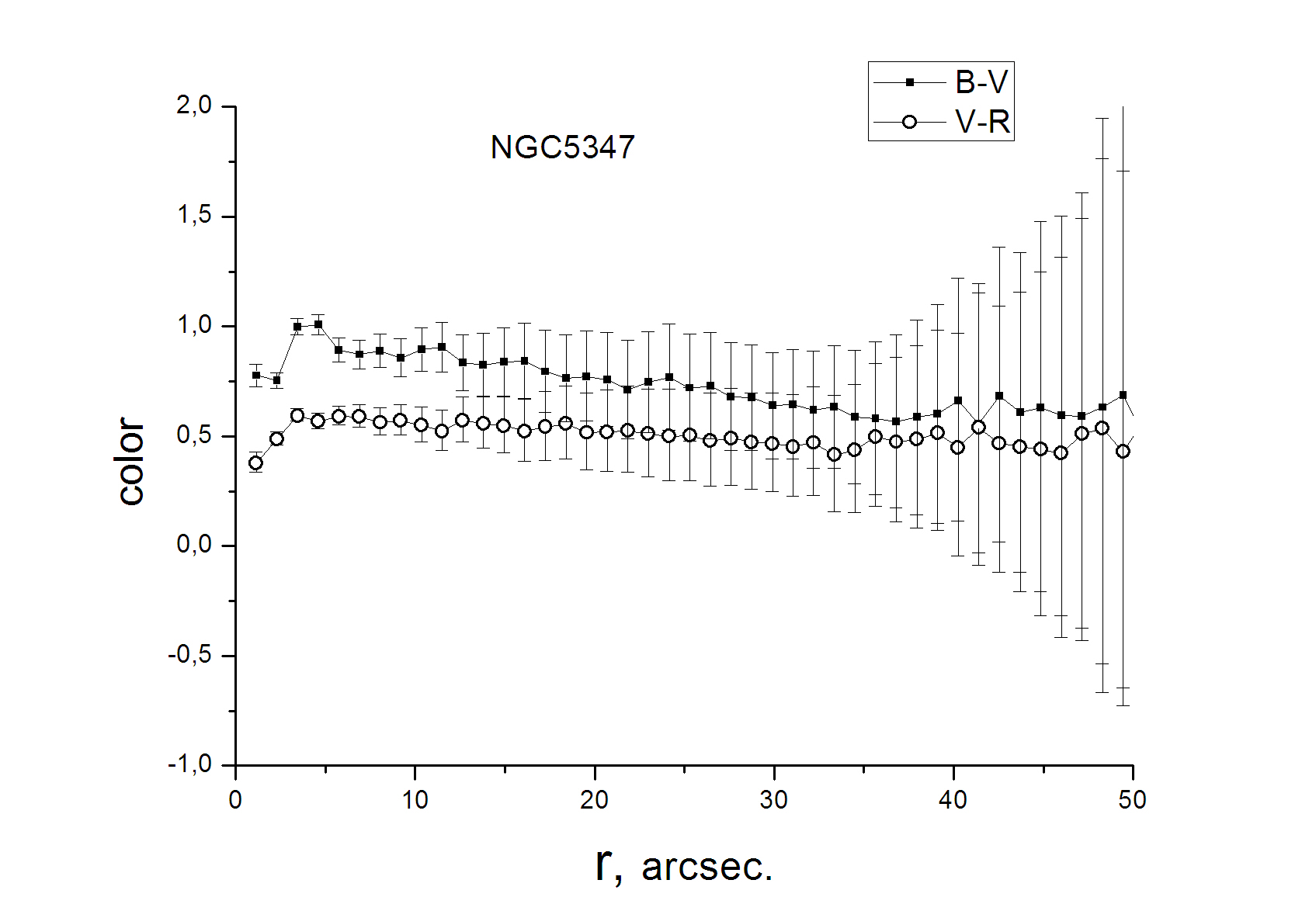}
\includegraphics[width=6.7cm,keepaspectratio]{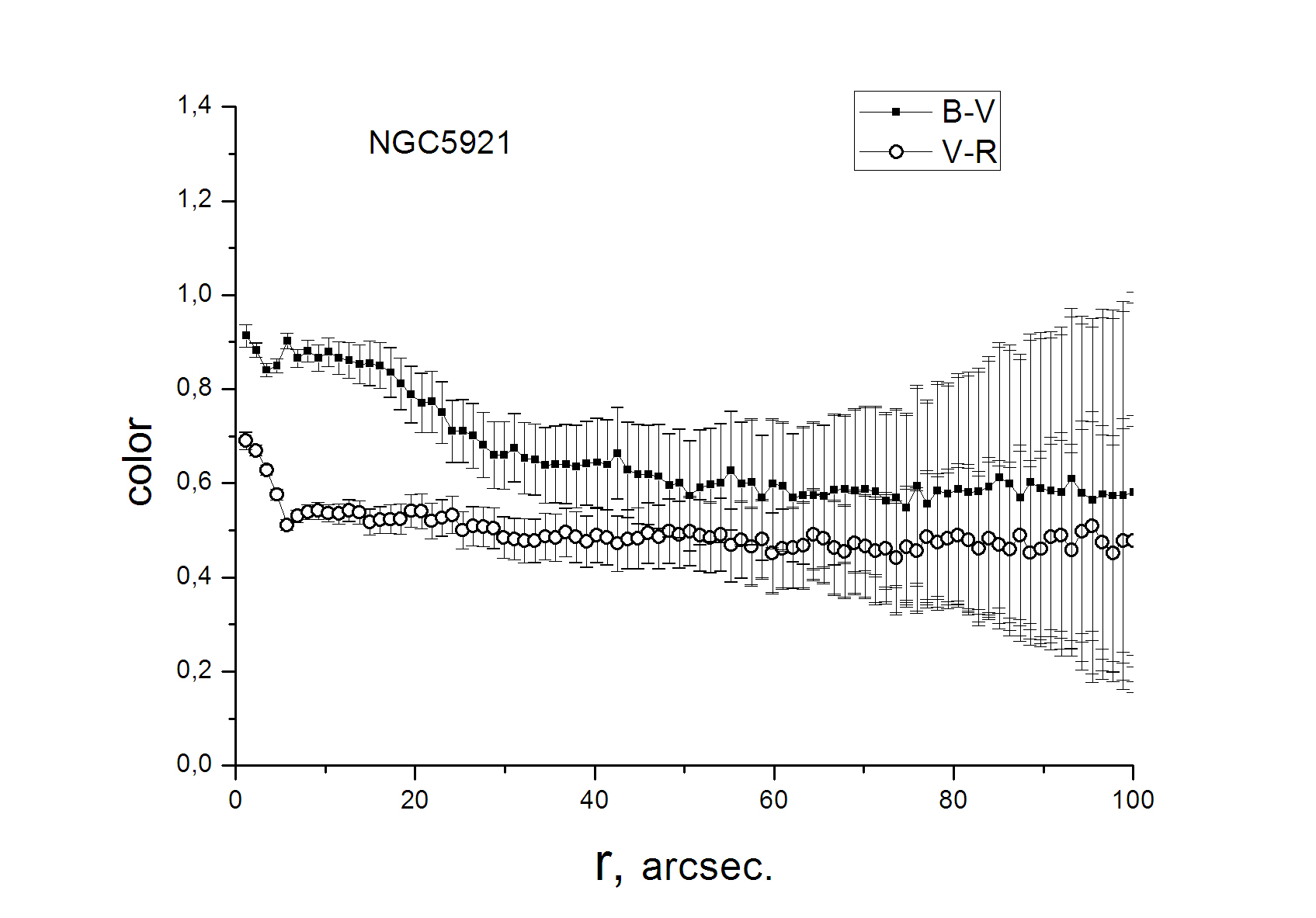}
\includegraphics[width=6.7cm,keepaspectratio]{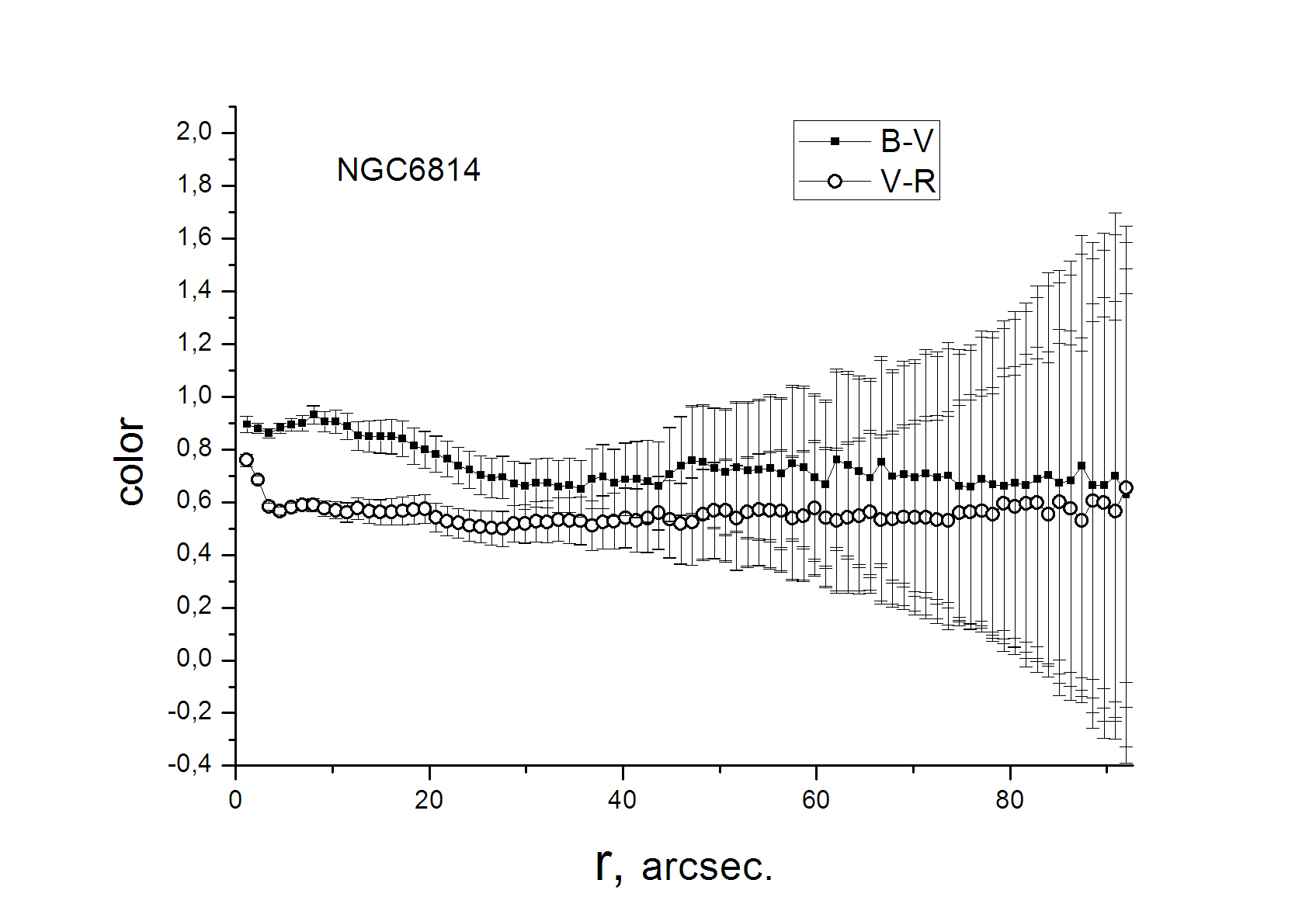}
\includegraphics[width=6.7cm,keepaspectratio]{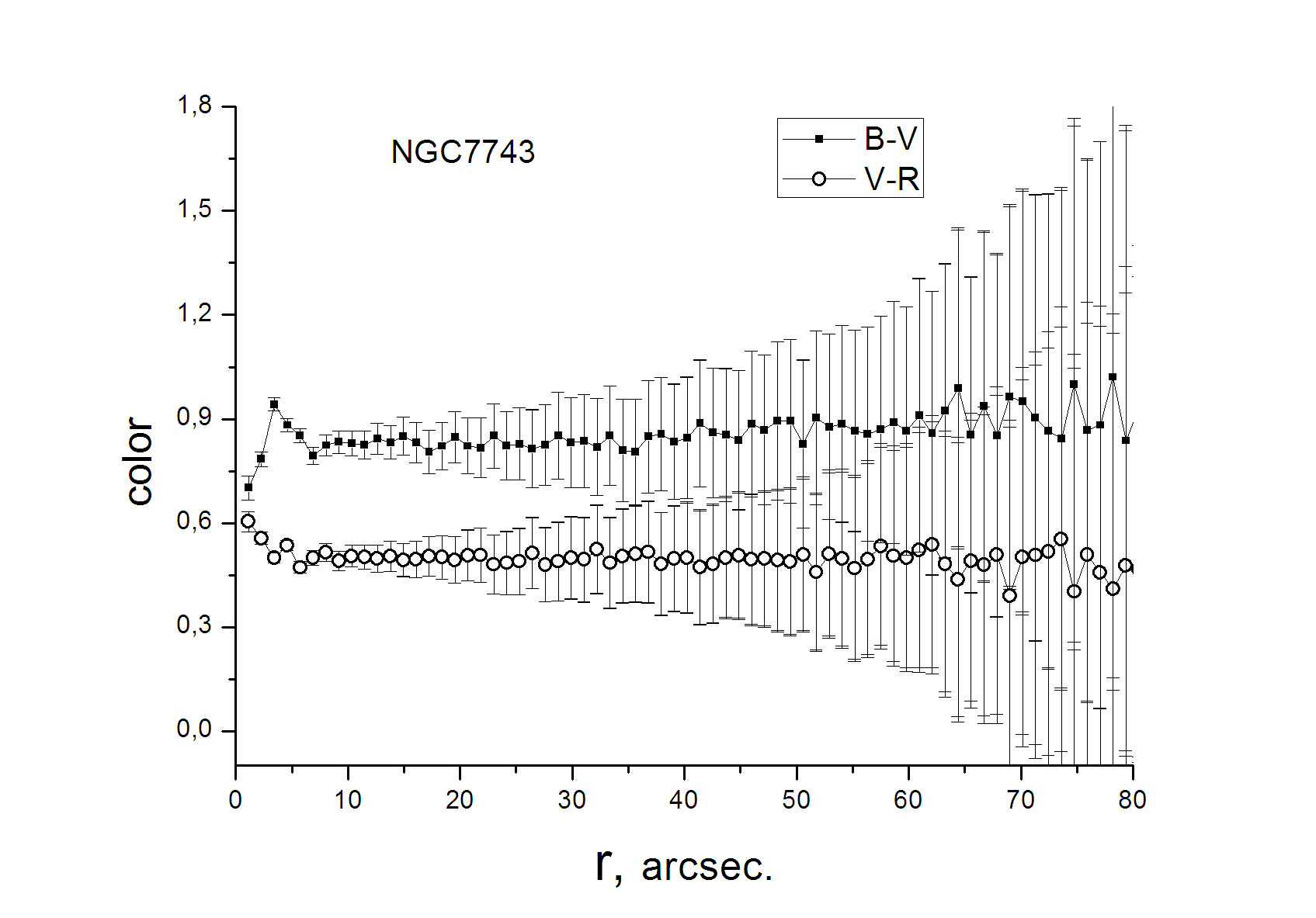}
\includegraphics[width=6.7cm,keepaspectratio]{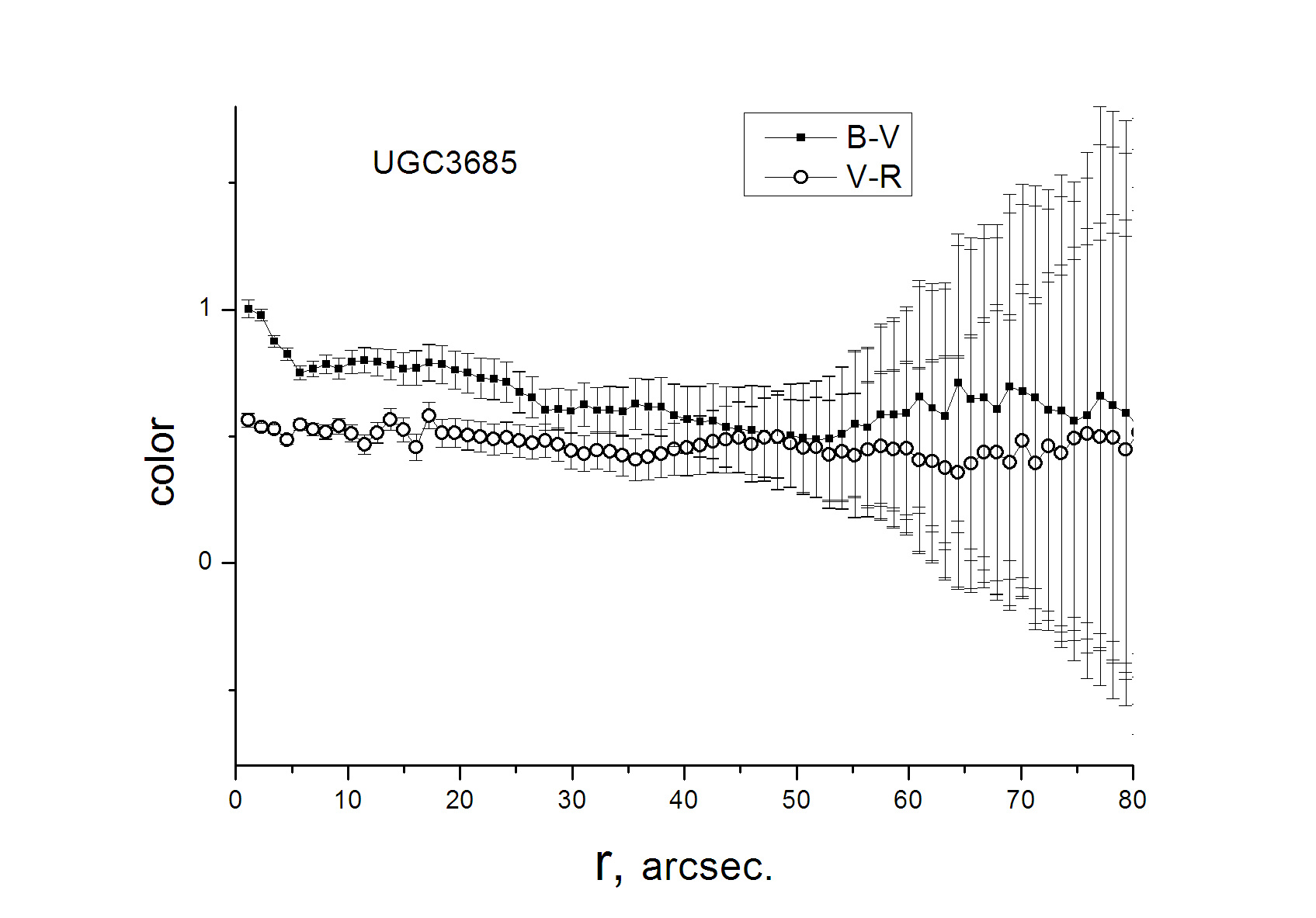}
\caption{Radial distributions of the azimuthally averaged colors.}
\end{figure}
\begin{figure}

\includegraphics[width=8cm,keepaspectratio]{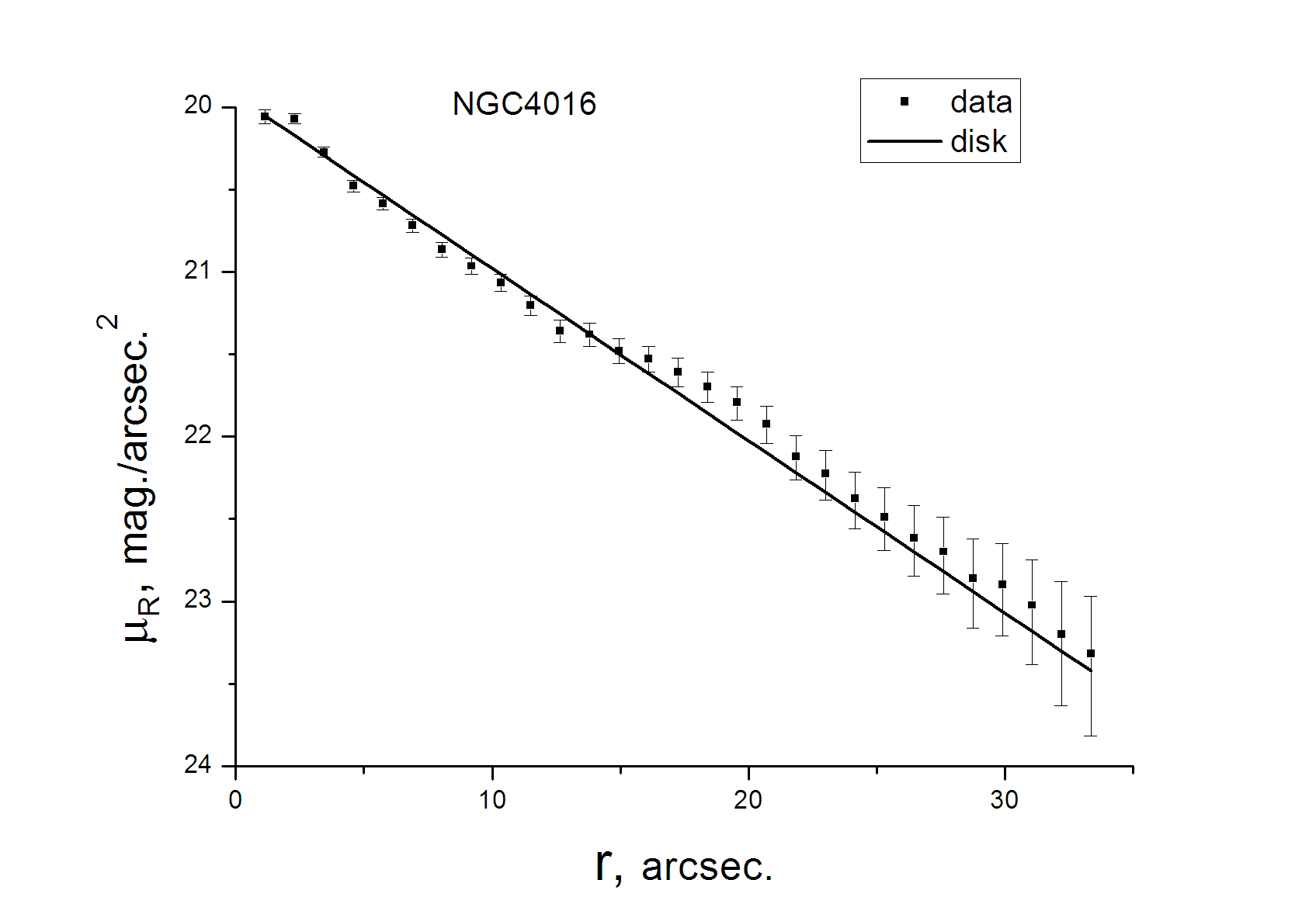}
\includegraphics[width=8cm,keepaspectratio]{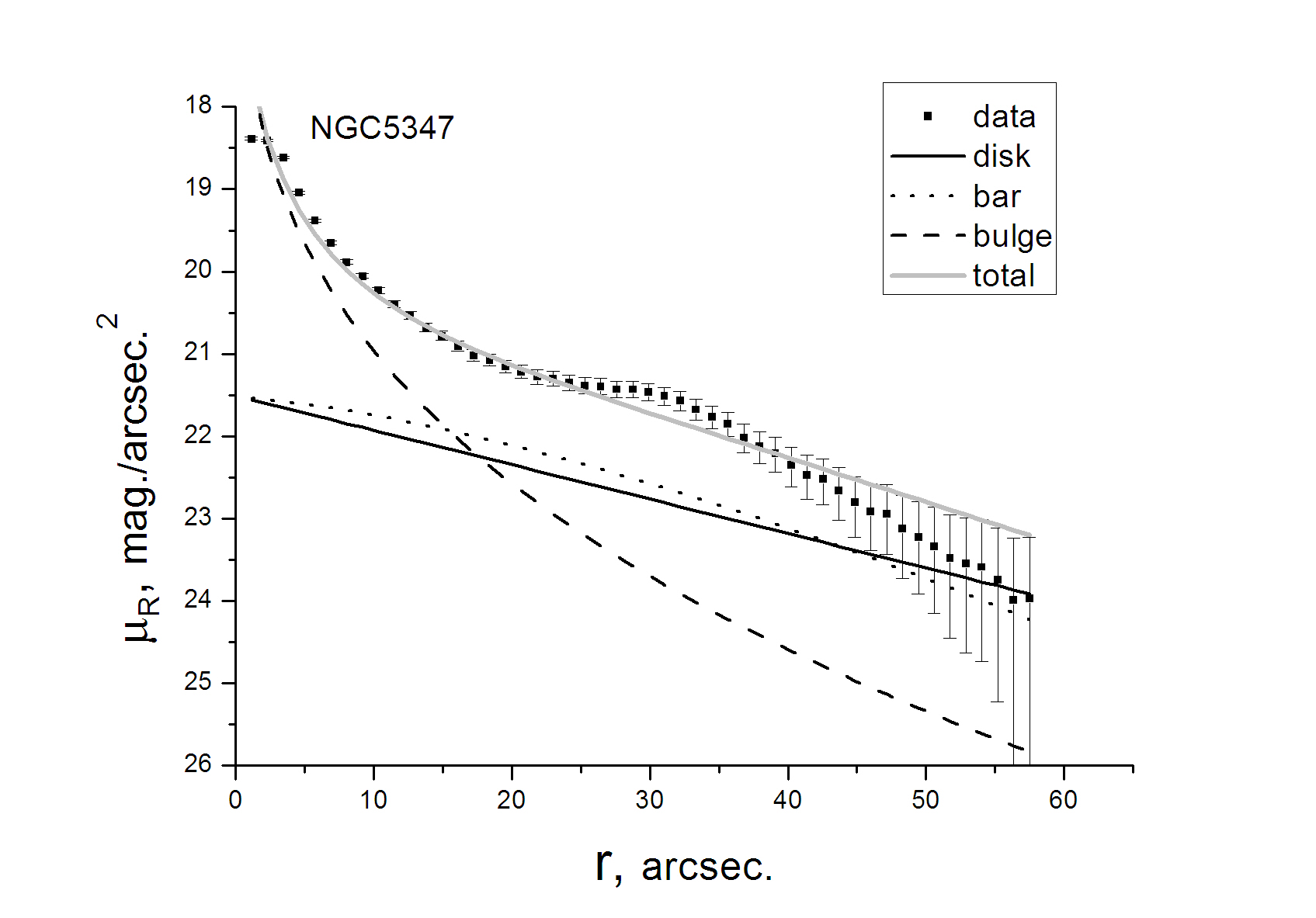}
\includegraphics[width=8cm,keepaspectratio]{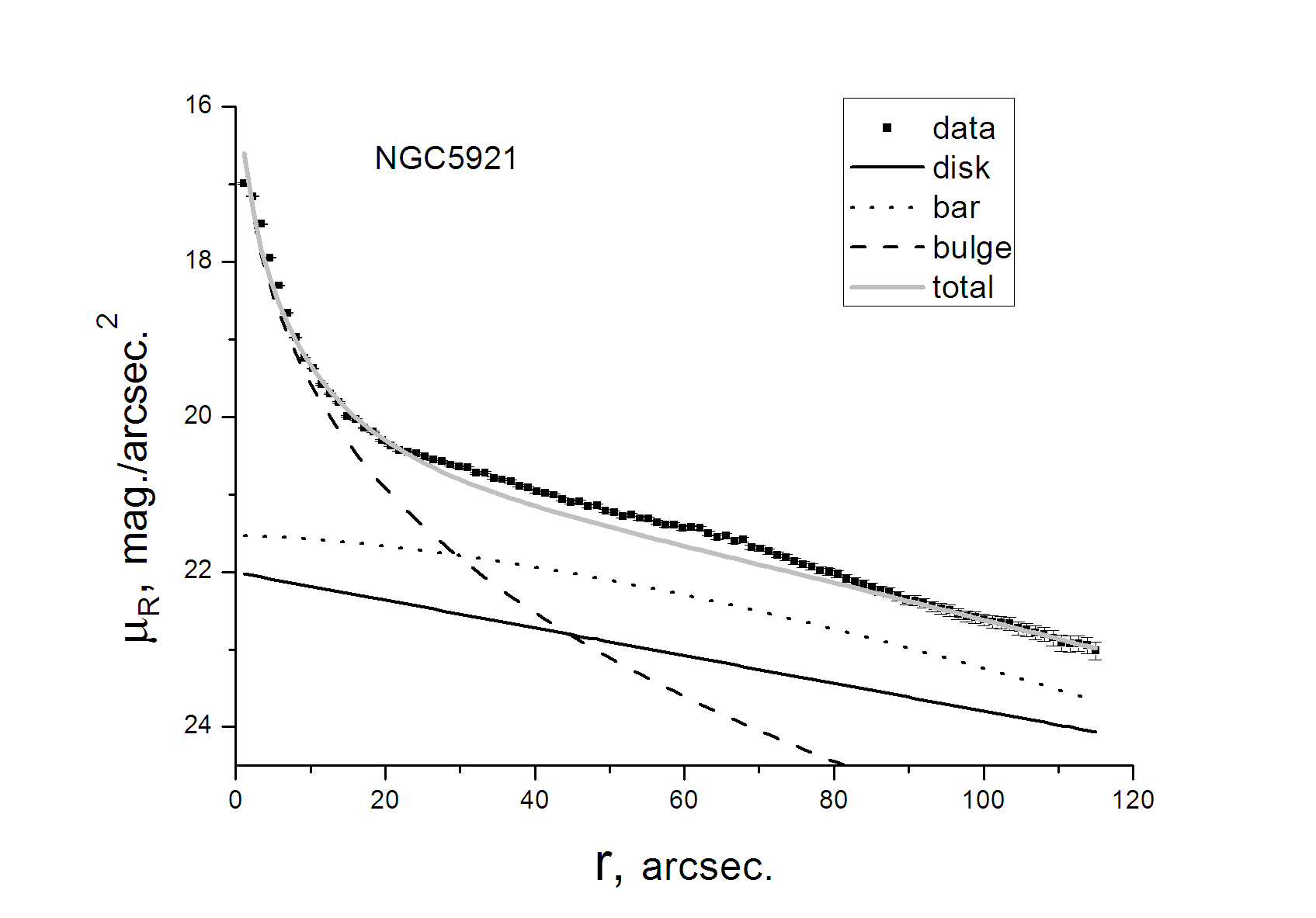}
\includegraphics[width=8cm,keepaspectratio]{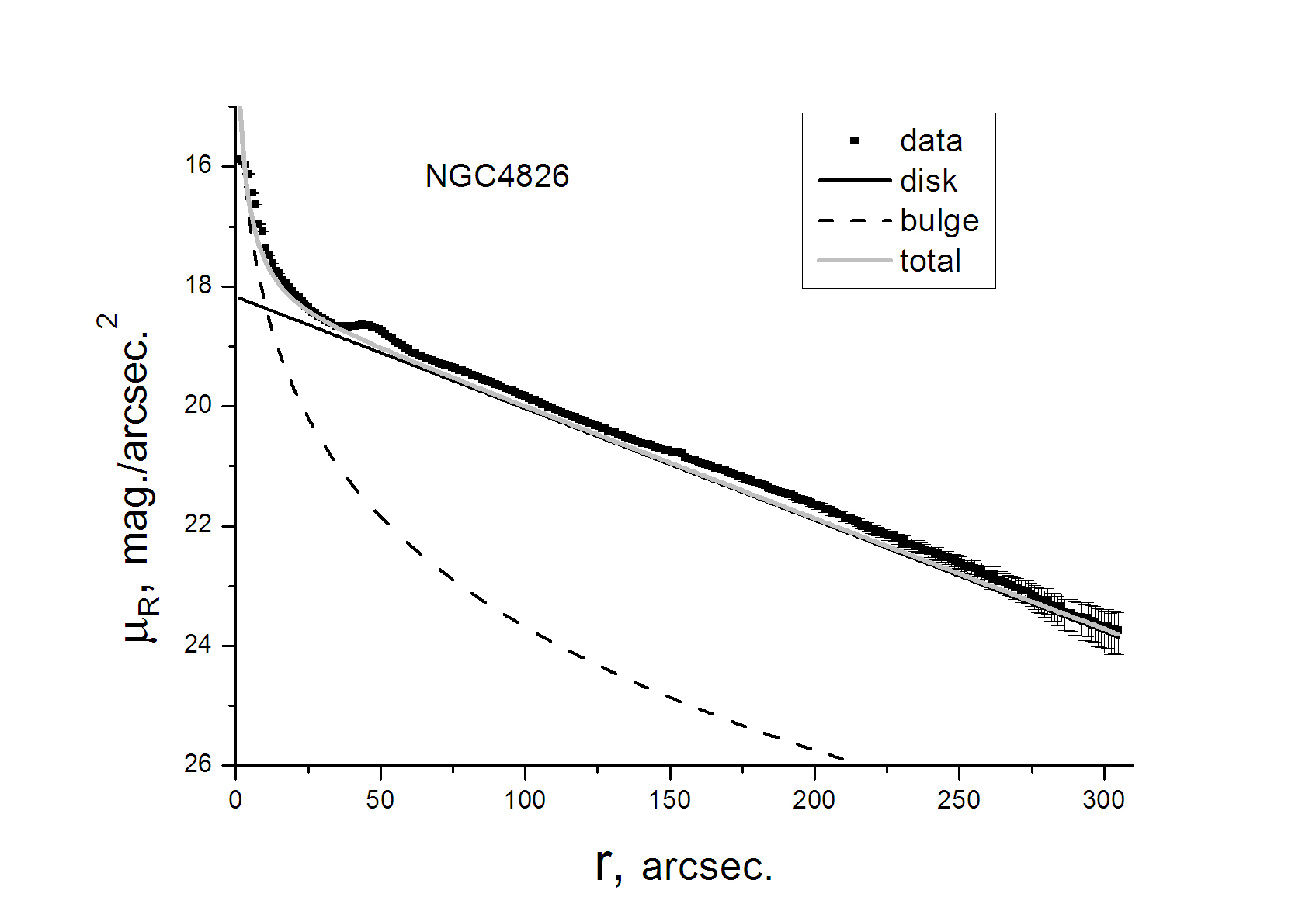}

\caption{The two-dimensional decomposition of the R-band surface brightness profiles 
into components (disk, bulge, and bar) using the BUDDA code. }
\end{figure}

The images were not successfully decomposed for two galaxies: NGC 1569 and NGC 4214, 
both are without noticeable bulges, because their disk brightness profiles
can be very poorly
described by an exponential law. We show no data on these galaxies in Table
\ref{table3}). Nevertheless their observed profiles were considered
when we photometrically estimated the disk contribution to the total rotation
curve.

\begin{figure}
\includegraphics[width=8cm,keepaspectratio]{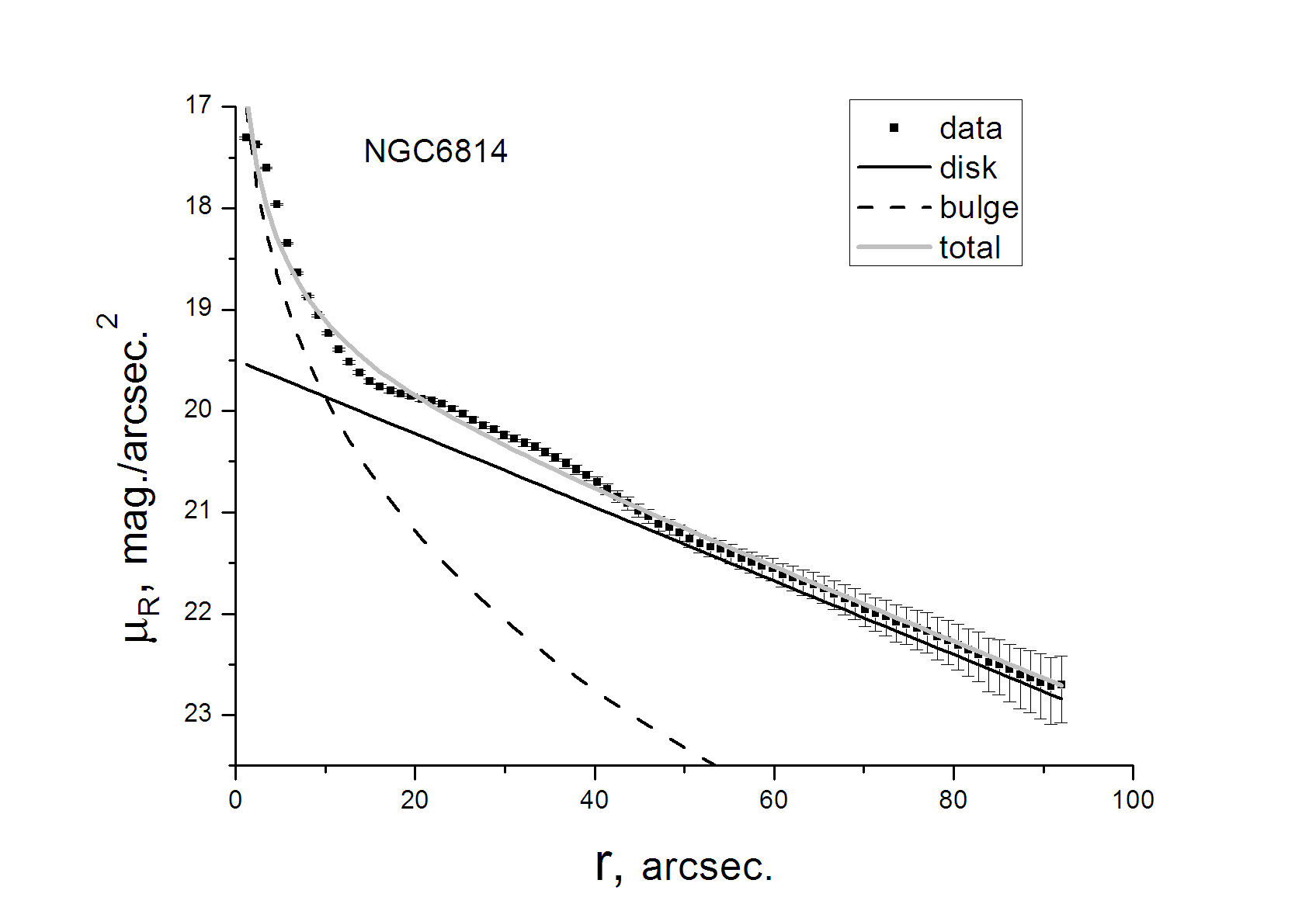}
\includegraphics[width=8cm,keepaspectratio]{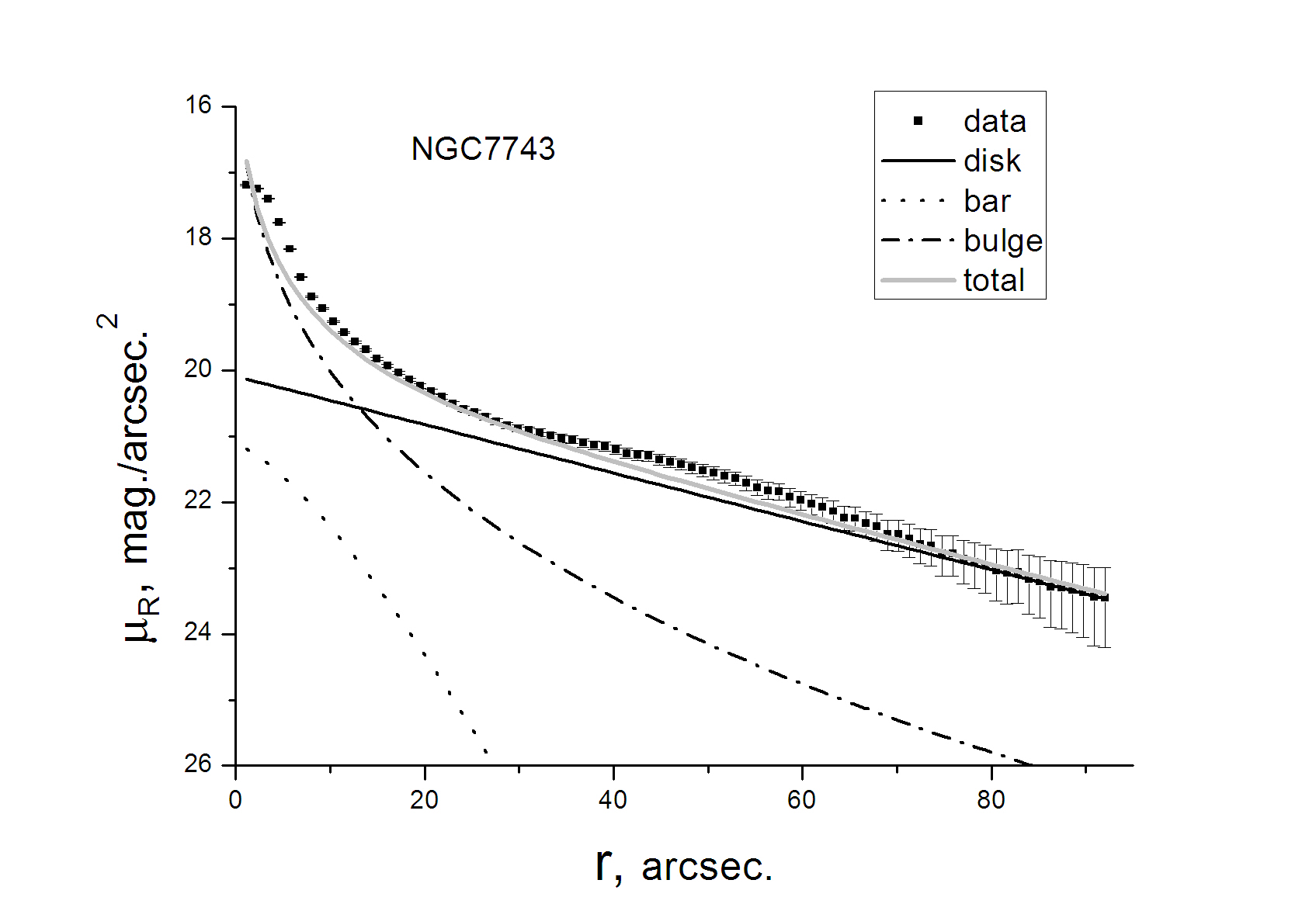}
\includegraphics[width=8cm,keepaspectratio]{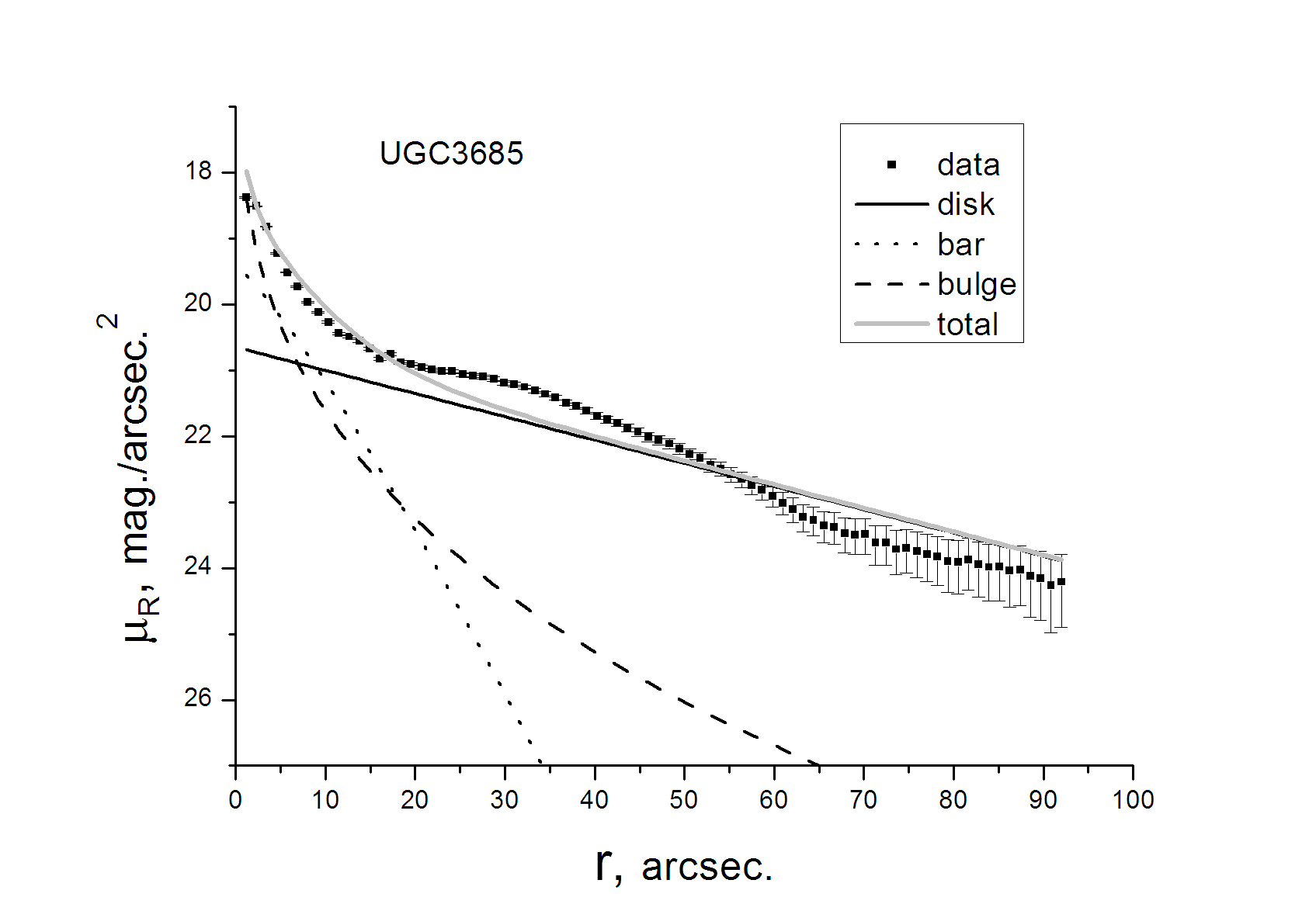}
\caption{Same as Fig. 4 for the galaxies UGC 3685, NGC 7743, and NGC 6814.}
\end{figure}

The quality of the image decomposition is clearly demonstrated in Fig. 6 for
NGC 5437. The image of this galaxy before and after the subtraction of the model
image generated by the BUDDA code is shown (the model contains disk, bulge and 
bar). Both images have the same contrast. Only a faint ring and regions of enhanced
brightness at the bar ends are noticeable in the residual image.

The BUDDA code reliably estimates the structural parameters of disks
(see
Figs. 4- 6, Table \ref{table4}), although, according to Gadotti (2008), the bulge-to-disk
luminosity ratio can be systematically overestimated by 5\% while the relative disk
luminosity can be systematically underestimated.
Below
we also consider the total $M/L$ ratios independent of the photometric decomposition
results, along with the mass-to-light ratios for the disks.
\begin{landscape}
{\scriptsize
\begin{table}
\caption{Structural parameters of the galaxies\label{table3}}

  %\begin{center}
    \begin{tabular}{|c|c|c|c|c|c|c|c|c|c|}
    \hline
{\scriptsize Galaxy}&$r_d$, {\scriptsize$''$}&$\mu_0$, {\scriptsize $^m/\Box''$}&$r_{e}$, {\scriptsize $''$}&$\mu_{e}$, {\scriptsize $^m/\Box''$}&$n_{bulge}$&$r_{e ~bar}$, {\scriptsize ''}&$\mu_{e ~bar}$, {\scriptsize $^m/\Box''$}&$n_{bar}$&{\scriptsize Band}\\
\hline
1&2&3&4&5&6&7&8&9&10\\
\hline
{\scriptsize NGC4016}   &   {\scriptsize 10.2   $\pm    0.27$}  &   {\scriptsize  20.6  $\pm     0.05$}  &               &               &               &              &               &                &B \\
    &   {\scriptsize  10.35 $\pm    0.27$}  &   {\scriptsize 20.3   $\pm     0.05$} &            &               &               &              &               &                &V \\
    &   {\scriptsize  10.4  $\pm    0.25$}  &   {\scriptsize 20.0   $\pm     0.04$} &            &               &               &              &               &                &R \\
\hline

{\scriptsize NGC4826}   &   {\scriptsize  59.8  $\pm    2.51$}  &   {\scriptsize 19.7   $\pm     0.08$} &    {\scriptsize 11.0  $\pm    0.50$}  &    {\scriptsize  20.3  $\pm   0.1$}   &    {\scriptsize 4.33  $\pm 0.49$}     &               &               &                &B \\
    &   {\scriptsize 59.2   $\pm    1.91$}  &   {\scriptsize 18.8   $\pm     0.07$} &    {\scriptsize 12.2  $\pm    0.55$}  &    {\scriptsize 19.5   $\pm    0.08$} &    {\scriptsize 5.05  $\pm 0.48$}     &               &               &                &V \\
    &   {\scriptsize 58.2   $\pm    1.42$}  &   {\scriptsize 18.3   $\pm     0.05$} &    {\scriptsize 16.7  $\pm    0.59$}  &    {\scriptsize 19.4   $\pm   0.07$}  &    {\scriptsize 6.3   $\pm 0.51$  }   &               &               &               &R   \\
\hline

{\scriptsize NGC5347}   &   {\scriptsize 25.4   $\pm    3.84$}  &   {\scriptsize 22.5   $\pm     0.2$}  &    {\scriptsize 5.74  $\pm    0.29$}  &   {\scriptsize  21.7   $\pm    0.07$} &    {\scriptsize 2.61  $\pm    0.18$}  &   {\scriptsize 35.2   $\pm    5.56$}  &   {\scriptsize 24.3    $\pm    0.07$} &   {\scriptsize  0.693 $\pm    0.47$ }&     B  \\
    &   {\scriptsize 25.0   $\pm    3.35$}  &   {\scriptsize 21.9   $\pm     0.20$} &    {\scriptsize 7.51  $\pm    0.46$}  &    {\scriptsize 21.1   $\pm    0.11$} &    {\scriptsize 4.45  $\pm    1.31$}  &   {\scriptsize 32.9   $\pm    4.01$}  &   {\scriptsize 23.3    $\pm    0.11$} &   {\scriptsize  0.68  $\pm    0.33$}& V    \\
    &   {\scriptsize 26.0   $\pm    2.3$}   &   {\scriptsize 21.6   $\pm     0.15$} &    {\scriptsize 9.61  $\pm    0.43$}  &    {\scriptsize 20.9   $\pm    0.07$} &    {\scriptsize 3.15  $\pm    0.21$}  &   {\scriptsize 31.8   $\pm    2.47$}  &   {\scriptsize 22.7    $\pm    0.07$} &    {\scriptsize 0.69  $\pm     0.18$}&    R   \\
\hline

{\scriptsize NGC6814}   &   {\scriptsize 30.3   $\pm    2.87$}  &   {\scriptsize 21.5   $\pm     0.15$} &    {\scriptsize 8.92  $\pm    0.89$}  &    {\scriptsize 22.3   $\pm   0.11$}  &    {\scriptsize 3.46  $\pm    0.19$}  &               &               &       &B   \\
    &   {\scriptsize 29.6   $\pm    1.88$}  &   {\scriptsize 20.5   $\pm     0.10$} &    {\scriptsize 12.5  $\pm    1.03$}  &    {\scriptsize 21.8   $\pm   0.1$}   &    {\scriptsize 3.52  $\pm    0.17$}  &               &               &       &V   \\
    &   {\scriptsize 29.9   $\pm    1.21$}  &   {\scriptsize 20.0   $\pm     0.08$} &    {\scriptsize 20.3  $\pm    1.10$}  &    {\scriptsize 21.7   $\pm    0.07$} &    {\scriptsize 3.85  $\pm    0.11$}  &               &               &       &R   \\
\hline

{\scriptsize NGC7743}   &   {\scriptsize 25.8   $\pm    4.05$}  &   {\scriptsize 21.2   $\pm     0.21$} &    {\scriptsize 6.81  $\pm    0.56$}  &    {\scriptsize 21.2   $\pm    0.1$}  &    {\scriptsize 2.8   $\pm    0.09$}  &   {\scriptsize 5.56   $\pm    0.46$}  &   {\scriptsize 21.8    $\pm    0.09$} &   {\scriptsize 0.69    $\pm   0.28$}   &B \\
    &   {\scriptsize 28.6   $\pm    3.12$}  &   {\scriptsize 20.7   $\pm     0.17$} &    {\scriptsize 10.4  $\pm    0.95$}  &    {\scriptsize 20.9   $\pm    0.11$} &    {\scriptsize 3.02  $\pm    0.16$}  &   {\scriptsize 3.54   $\pm    0.1$}   &   {\scriptsize 20.9    $\pm    0.11$} &   {\scriptsize  0.75  $\pm    0.06$}  &V   \\
    &   {\scriptsize 29.5   $\pm    2.46$}  &   {\scriptsize 20.3   $\pm     0.13$} &    {\scriptsize 11.3  $\pm    0.77$}  &    {\scriptsize 20.5   $\pm    0.08$} &    {\scriptsize 3.24  $\pm    0.12$}  &   {\scriptsize 9.99   $\pm    2.33$}  &   {\scriptsize 22.5    $\pm    0.08$} &    {\scriptsize 0.737 $\pm    0.52$}   &R \\
\hline

{\scriptsize NGC5921}   &   {\scriptsize 43.6   $\pm    9.54$}  &   {\scriptsize 23.4   $\pm     0.29$} &    {\scriptsize 11.1  $\pm    0.65$}  &    {\scriptsize 21.4   $\pm   0.12$}  &    {\scriptsize 3.65  $\pm    0.04$}  &   {\scriptsize 77.1   $\pm    6.71$}  &   {\scriptsize 23.5    $\pm    0.12$} &   {\scriptsize  0.68  $\pm     0.2$}  &B  \\
    &   {\scriptsize 55.6   $\pm    5.29$}  &   {\scriptsize 22.1   $\pm     0.1$}  &    {\scriptsize 14.7  $\pm    0.55$}  &    {\scriptsize 21.3   $\pm   0.06$}  &    {\scriptsize 3.28  $\pm    0.09$}  &   {\scriptsize 72.9   $\pm    6.08$}  &   {\scriptsize 23.4    $\pm    0.06$} &    {\scriptsize 0.68  $\pm     0.19$} &V  \\
    &   {\scriptsize 60.7   $\pm    3.9$}   &   {\scriptsize 22.1   $\pm     0.12$} &    {\scriptsize 18.54 $\pm    0.45$}  &   {\scriptsize  20.9   $\pm   0.03$}  &    {\scriptsize 3.93  $\pm    0.09$}  &   {\scriptsize 72.2   $\pm    2.91$}  &   {\scriptsize 22.7    $\pm    0.03$} &    {\scriptsize 0.64   $\pm   0.12$}  &R  \\
\hline

{\scriptsize UGC3685}   &   {\scriptsize 25.6   $\pm    1.91$}  &   {\scriptsize 21.5   $\pm     0.12$} &    {\scriptsize 5.34  $\pm    0.69$}  &    {\scriptsize 23.4   $\pm    0.25$} &    {\scriptsize 3.8   $\pm    2.45$}  &   {\scriptsize 3.34   $\pm    0.17$}  &   {\scriptsize 20.4   $\pm     0.25$} &   {\scriptsize 0.96    $\pm   0.26$}  &B   \\
    &   {\scriptsize 24.12  $\pm    1.3$}   &   {\scriptsize 20.8   $\pm     0.10$} &    {\scriptsize 6.95  $\pm    0.4$}   &    {\scriptsize 21.8   $\pm    0.11$} &    {\scriptsize 4.59  $\pm    0.99$}  &   {\scriptsize 4.14   $\pm    0.32$}  &   {\scriptsize 20.9    $\pm    0.11$} &   {\scriptsize 1.69    $\pm   0.69$}   &V \\
    &   {\scriptsize 30.8   $\pm    1.44$}  &   {\scriptsize 20.80  $\pm     0.05$} &    {\scriptsize 9.56  $\pm    0.38$}  &    {\scriptsize 21.68  $\pm   0.07$}  &    {\scriptsize 3.08  $\pm    0.29$}  &   {\scriptsize 8.62   $\pm    0.87$}  &   {\scriptsize 21.0 $\pm   0.07$} &   {\scriptsize 0.84   $\pm     0.12$}  &R \\
\hline
 \end{tabular}
\end{table}
}
\end{landscape}
\begin{figure}
\includegraphics[width=6.7cm,keepaspectratio]{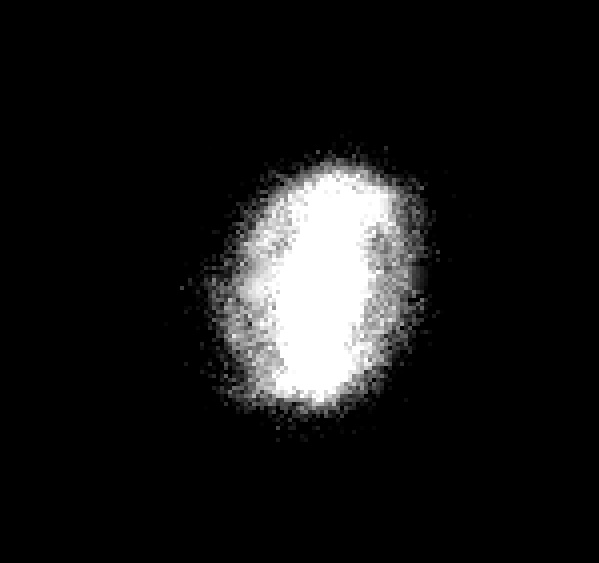}
\includegraphics[width=6.7cm,keepaspectratio]{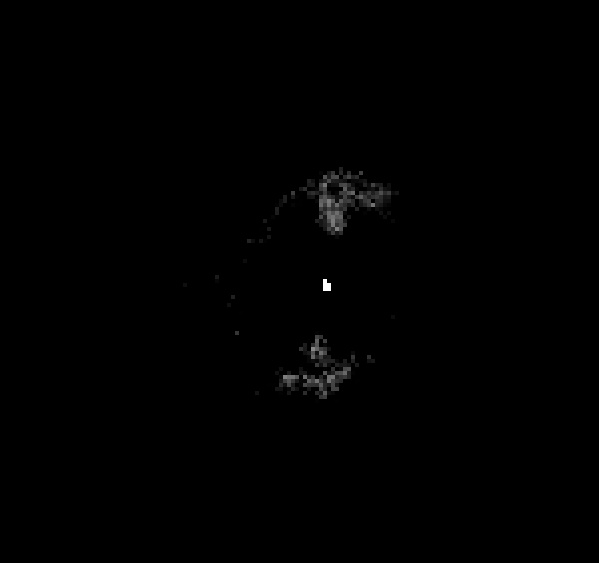}
\caption{The R-band image of the galaxy with a bright bar NGC 5347 before and after the 
subtraction of the model image generated
by the BUDDA. Both images have the same contrast.}
\end{figure}

\section{COMPARISON WITH OTHER PUBLICATIONS} 
We compare the surface brightness profiles with results
by other authors. In Fig. 7, the profiles of all galaxies (except NGC 1569)
were corrected for extinction in the Galaxy, but were not corrected for the disk
inclination. Fig. 7 shows good agreement between our estimates and
published results.
An exception is NGC 1569, which exhibits a significant discrepancy at the periphery
between our R-band brightness profile and those from Swaters \& Balcells (2002) and Stil
\& Israel (2002). This discrepancy may be explained by diferent procedures
we used  for subtraction of numerous foreground stars for this galaxy.
There is also a
noticeable discrepancy at the periphery between our R-band profile of NGC 6814 and that
from de Robertis et al. (1998). However, there is a good agreement with the
profile from S\'anchez-Portal et al. (2000) observed in the same band (see Fig. 7). In the
remaining cases, the difference between our profiles and those of other authors does not
exceed $0.3^m$. This is comparable to the typical errors in the surface brightness estimates far
from the center.

\begin{figure}
\begin{minipage}[h]{0.5\linewidth}
\includegraphics[width=7cm,keepaspectratio]{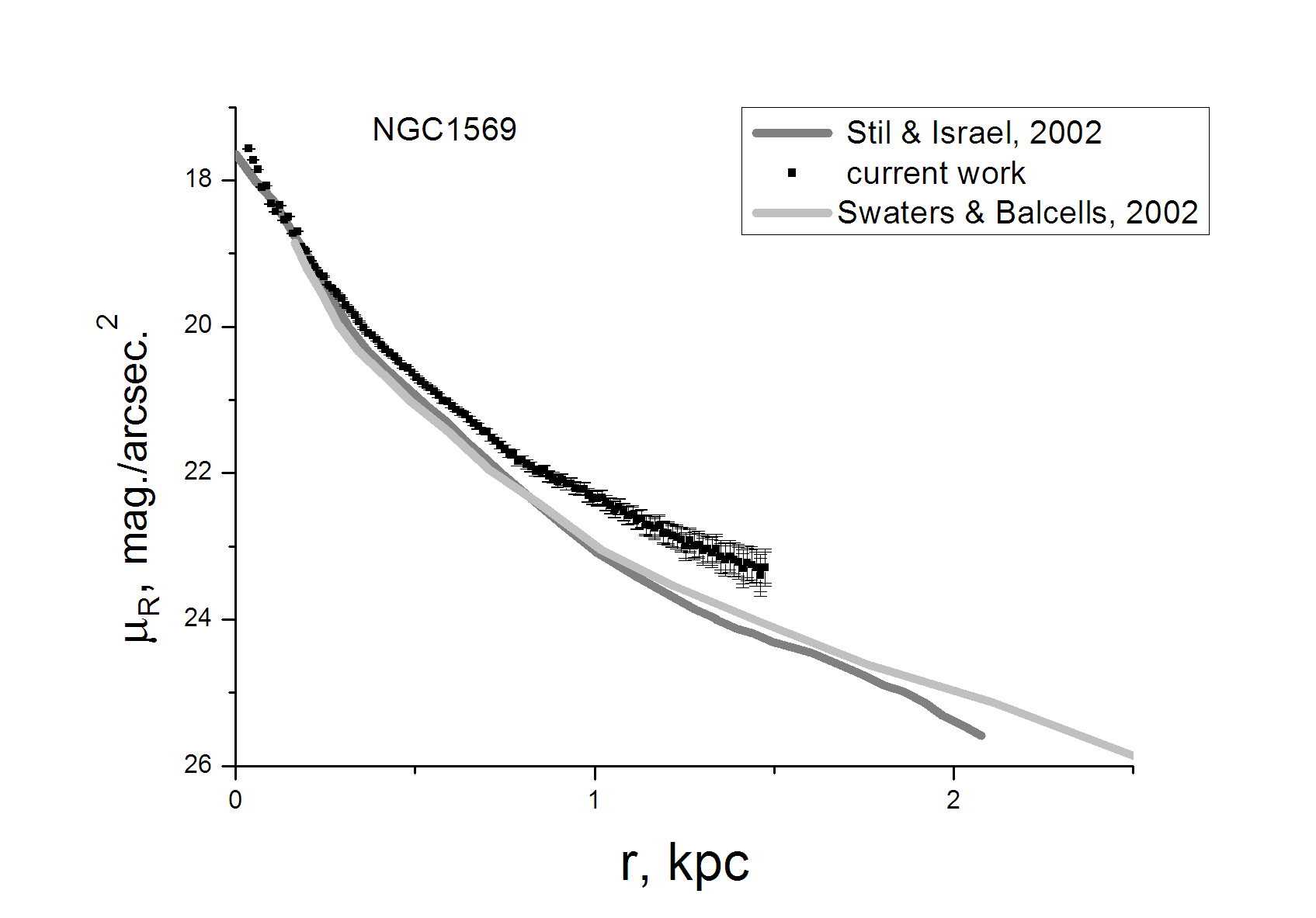}
\end{minipage}
\begin{minipage}[h]{0.5\linewidth}
\includegraphics[width=7cm,keepaspectratio]{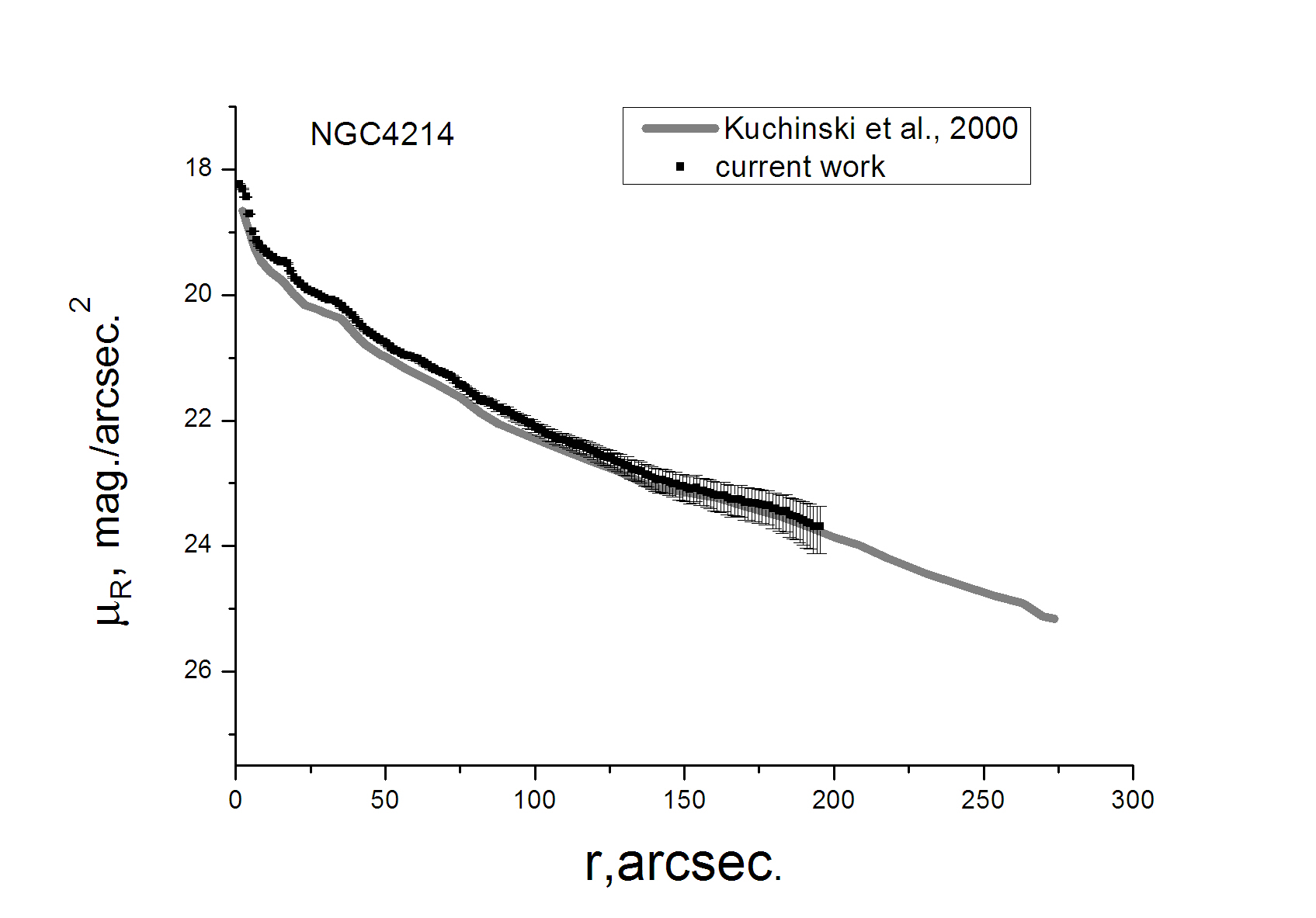}
\end{minipage}
\begin{minipage}[h]{0.5\linewidth}
\includegraphics[width=7cm,keepaspectratio]{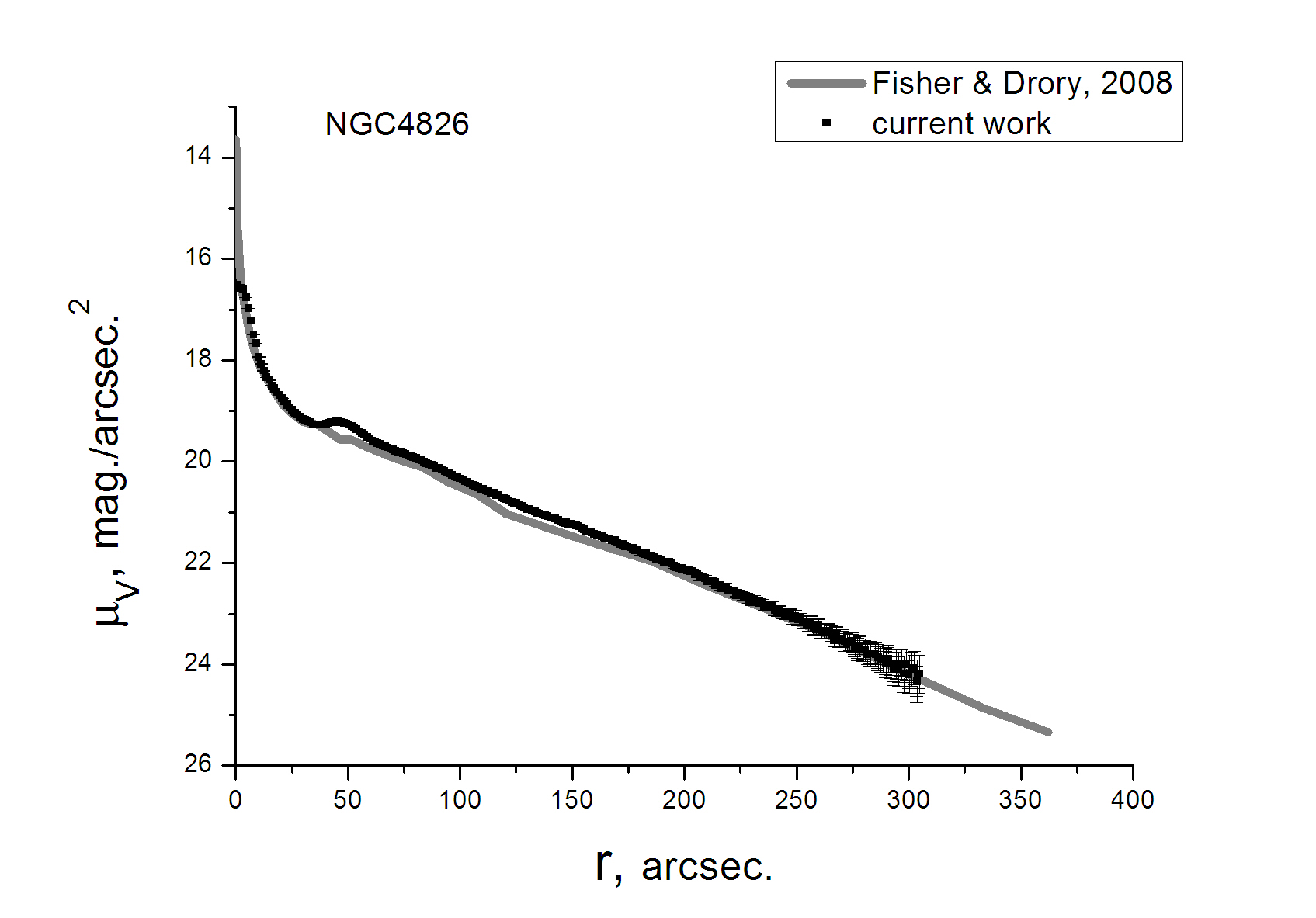}
\end{minipage}
\begin{minipage}[h]{0.5\linewidth}
\includegraphics[width=7cm,keepaspectratio]{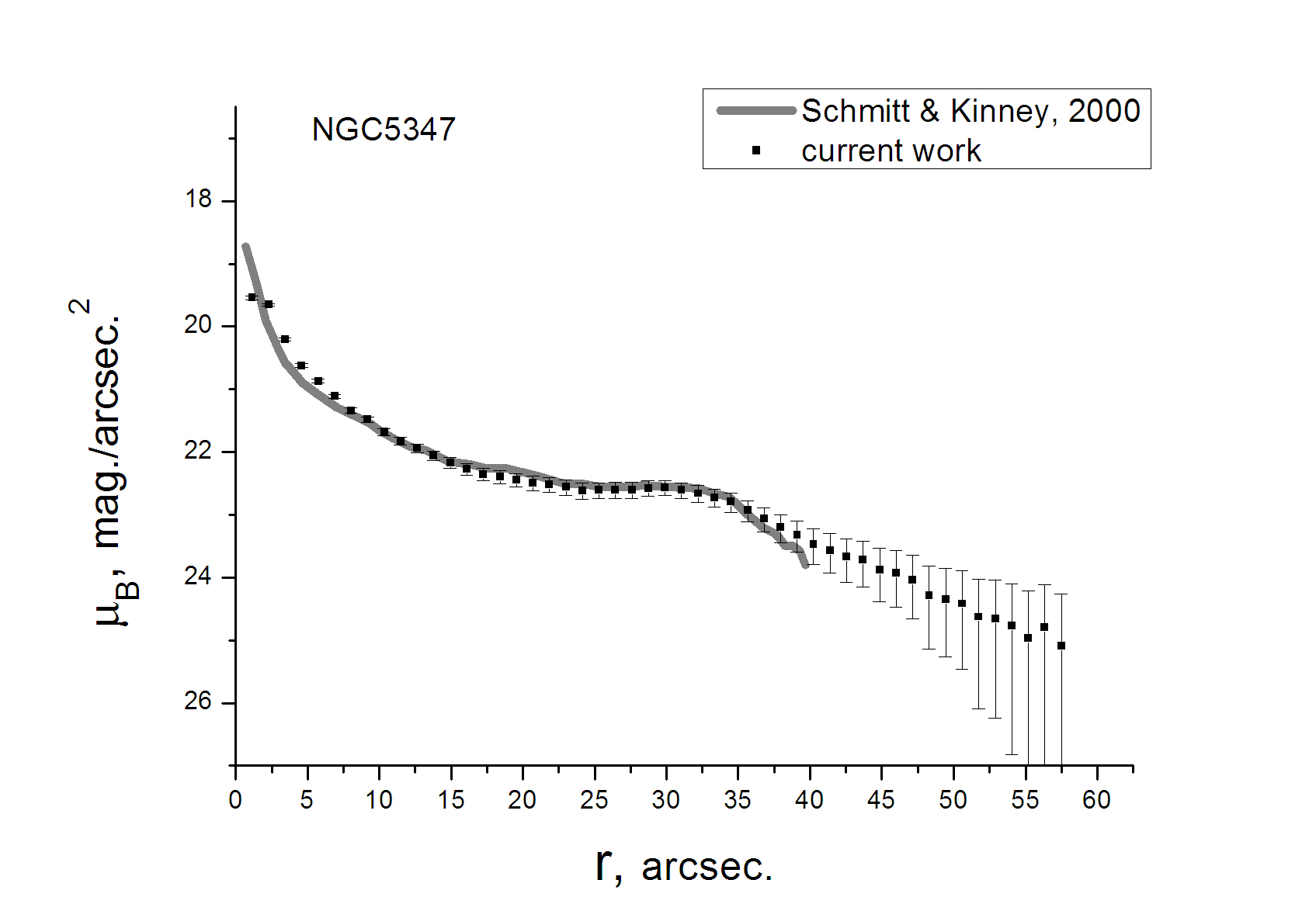}
\end{minipage}
\begin{minipage}[h]{0.5\linewidth}
\includegraphics[width=7cm,keepaspectratio]{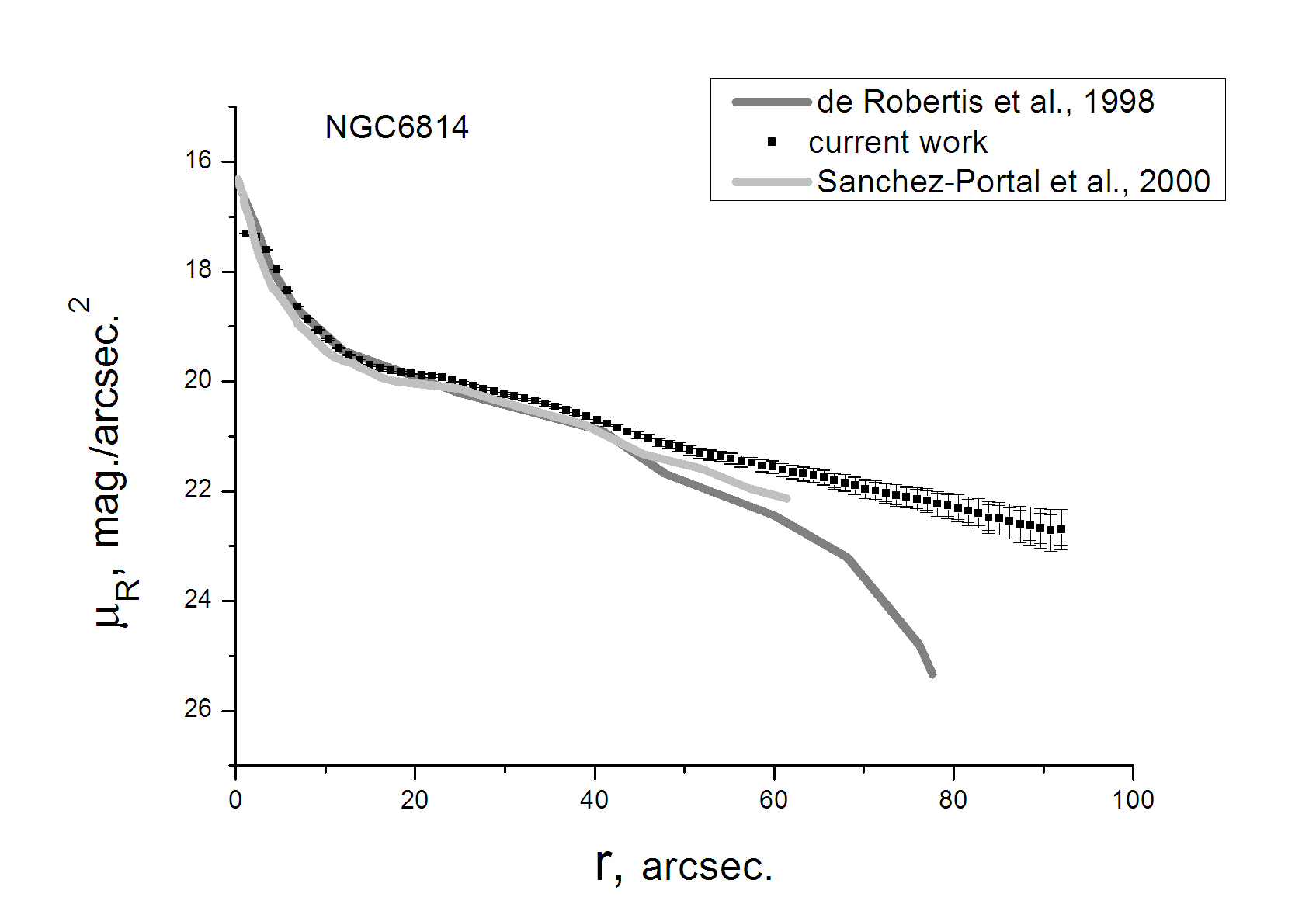}
\end{minipage}
\begin{minipage}[h]{0.5\linewidth}
\includegraphics[width=7cm,keepaspectratio]{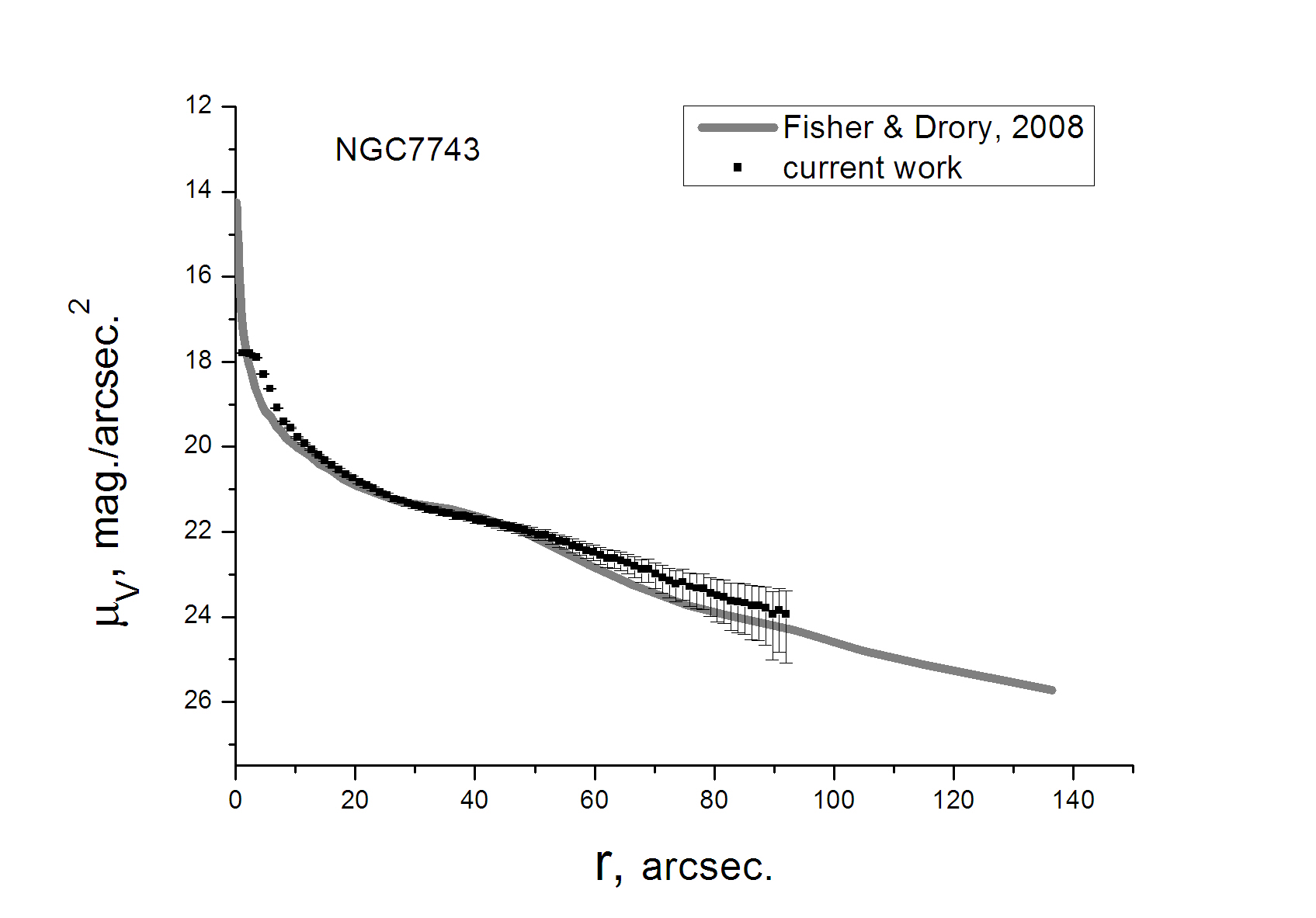}
\end{minipage}

\caption{Comparison of the $B,V,R$ brightness profiles with the results of other authors.}
\end{figure}

\section{COMPARISON OF THE PHOTOMETRIC AND DYNAMICAL DISK MASS ESTIMATES}
We choose two conditions that allow
us to suspect the existence of a discrepancy between
the photometric and dynamic mass estimates of the stellar population of a galaxy:\\
(i) The photometric $M/L$ of stellar population of a galaxy within some
radius (taken within four radial scale lengths $r_d$) obtained from the total  color
index, exceeds the dynamical estimate of $M/L$ determined from the relation
$(M/L)_{dyn}=4v_{max}^2r_d/(GL_{tot})$, where $L_{tot}$ is the total luminosity
(including bulge) and $v_{max}$ is
the maximum rotational velocity of the galaxy.\\
(ii) The maximum rotational velocity of the disk component
$v_{max~disk}$ estimated from the photometric
disk mass and scale length exceeds the observed maximum
rotational velocity of the galaxy $v_{max}$.
The value of $v_{max~disk}$ can be determined from
the radial surface density profile $\sigma(R)$ obtained from the
brightness distribution and the model $M/L$ ratio estimated from
the measured color index. The rotation curve in spiral galaxies
is usually a constant beyond $r=(1.5-2)r_d$. 
Hence we assume that
$v_{max}$ is close to the rotation velocity of galaxy at
$r\approx 2r_d$, where the disk component of rotation curve approaches
$v_{max~disk}$. 

For all galaxies except NGC 4214 and NGC 1569 the disk $M/L$
ratio was assumed to be constant along the radius, while
the radial change of $M/L$ in the two mentioned objects was
determined from the color profiles. For an exponential disk density
distribution adopted for all galaxies except NGC 4214 and NGC 1569,
$v_{max~disk}\approx 0.623(GM_d/r_d)^{0.5}$, where $M_d=2\pi
G\sigma_0r_d^2$ is the disk
mass and $\sigma_0$ is the central surface density of the disk.\\

Galaxies with abnormally low $M/L$, which is at least a factor of 3 lower than the
expected ratio for a given $(B-V)_0$ color, are rare - they account for
about 3\% of the number of galaxies of S0 and later types which have the measured
rotational velocities (from the H\,{\small I} line width) and $(B-V)_0$  colors (see Fig. 1 in
Saburova et al. 2009). In fact, their number may be considerably smaller.
There are two factors which may be responsible for the discrepancy
between the
dynamical and photometric disk mass estimates:\\
(i) errors in the estimates of observed quantities,
such as the circular velocity, the inclination, the distance
to the galaxy, the luminosity and color;\\
(ii) inconsistency of the galaxy's stellar population with  the evolutionary population
synthesis model used for the $M/L$ estimate: the ignorance of starburst,
low metallicity Z of the stellar population, and/or an anomalous stellar IMF. Below we
consider each of these factors in our galaxies.\\

Note that all galaxies of our sample except NGC 1569 and
NGC 4016 follow the sequence of $(B-V)-(V-R)$ colors for
normal galaxies derived by Buta \& Williams  (1995). This suggests
the absence of significant peculiarities in the star formation history
and confirms the applicability of population
synthesis models (Bell and de Jong 2001). 
The star formation in NGC 1569 is very active, then we apply the $M/L$
--- color relation including a starburst. The second galaxy,  NGC 4016, was included in the
sample due to the unrealistically low $(B-V)_0$ given by NED, but it is
not confirmed in our work, so we conclude that this is not a galaxy with an 
abnormally low $M/L$ (see below).\\

{\bf NGC1569} is a Magellanic-type irregular dwarf starburst galaxy.
Active star formation in this object is probably associated with
tidal interaction (Grocholski et al. 2008). According to Recchi et
al. (2006), the chemical abundance in this galaxy is consistent with
the assumption that there were long-lasting episodes of quiet star
formation completed by series of starbursts. Since NGC 1569 lies
near the Galactic plane, the extinction correction is
very uncertain for it: the correction coefficients in Schlegel et
al. (1998) and Burstein and Heiles (1982) for the $B$ band differ by
$1^m$. This galaxy is a probable cause of the low mass-to-light ratio
estimate. Therefore, we can adopt the photometric $Ks$-band profile from
Vaduvescu et al. (2006), along with the $B, V, R$ 
magnitudes, in order to estimate the disk mass from the photometry
and its contribution to the rotation curve. Using the $Ks$ band
also allows us to reduce the effects of intense star formation on the
model estimate of the mass-to-light
ratio.\\

The ionized gas velocity field of NGC 1569 obtained within the framework of the GHASP
project (Epinat et al. 2008) covers the central parts of the galaxy and shows no evidence
of rotation. However, the rotation curve determined by Stil and Israel (2002) in the
H\,{\small I} line allows the maximum rotational velocity $v_{max}$ to be estimated. We
compared $v_{max}$ with the maximum rotational velocity of the disk component of the
rotation curve $v_{max~disk}$ by estimating the disk mass from photometric data. The disk
density profile needed for this purpose was calculated from the $Ks$ and $R$ brightness
profiles as well as from the $(V -R)$ color distribution and the total $ (J-K)$ color
obtained by Vaduvescu et al. (2006). We determined the expected $M/L_R$ and
$M/L_K$ ratios for the disk (see Table \ref{table4}) from the color indices
following the models of Bell and de Jong (2001) calculated for a
modified Salpeter IMF with the inclusion of  a starburst.\footnote {We
took the starburst duration 0.5 Gyr and the fraction of young stars is 10\% of the total
mass of the stellar population.} Taking into account the radial brightness profiles in these
photometric bands, we obtained the radial disk surface density profile and the
corresponding values of $v_{max~disk}$ (see Table \ref{table4}). It follows that
$v_{max~disk}$ determined from $Ks$ photometry turns out to be close to the observed
maximum rotational velocity. Hence we may conclude that this object does not have an
abnormally low $M/L$. Note that the mass of its dark halo within the optical boundaries
should be small compared to the disk mass, as suggested by the
closeness of the $v_{max~disk}$ and $v_{max}$ estimates.\\

{\bf NGC4016} is an SBd-type galaxy. We included it in our sample
due to its unrealistically low color index $(B-V)=0.03$  in the NED,
which we do not confirm (see Table \ref{table2}). The rotational
velocity $v_{max~disk}$ determined from the photometric density
estimate using our color and luminosity measurements turns out to be
lower than the total rotational velocity $v_{max}$ (see Table
\ref{table4}). This gives no grounds for attributing
it to objects with anomalous $M/L$.\\

{\bf NGC4214}, just as NGC 1569, is a Magellanic type irregular galaxy with a bar. The
main problem related to the mass estimation for this galaxy is an unreliably determined
inclination (see Fig. 2). The axial ratio of NGC 4214 changes along the radius 
from 0.7 to 0.94, which corresponds to the inclinations $i_1=45^o$ and $i_2=20^o$. The
rotational velocity $v_{max~disk}$ found from the  R-band brightness profile and
the $(B - R)$ and  $(B - V )$ colors for a modified Salpeter IMF, including a
starburst similar to that adopted for NGC 1569, is compared in Table \ref{table4} with the
observed rotational velocity $v_{max}$ of the galaxy for both inclination angles. As follows from our estimates, $v_{max}$ may be either higher or
lower than $v_{max~disk}$ if to choose  $i_2=20^o$ or $i_1=45^o$, respectively.
Obviously, the observed flattening of the galaxy's inner part is distorted by the presence
of a bar, so  the peripheral values of $b/a$ are preferable for the inclination
estimate. However, the contradiction between the photometric and dynamical models is
retained even at $i_2=20^o$:
 in this case, in spite of
$v_{max}>v_{max~disk}$, the radius at which the disk component of
the rotation curve, obtained from photometry data,  has a
maximum ($R \approx 2$ kpc) turns out to be a factor of 3 smaller
than the radius where the observed rotation curve derived by Allsopp
(1979) reaches its maximum. Thus, the disk with the photometrically
calculated density distribution is inconsistent with the observed
rotation curve. This conclusion remains valid even if we take into
account the low linear resolution of the H\,{\small I} data in
Allsopp (1979) ($\Delta R \approx 2$ kpc). However, there
remains the possibility that the discrepancy between $v_{max~disk}$
and the rotation velocity  in the  central part of the
galaxy can be associated with noncircular motions of gas due to the
presence of a bar. Therefore, although an anomalous stellar
composition in NGC 4214 remains quite possible, the available data
do not allow a reliable $M/L$ estimate to be obtained for this
galaxy. A more detailed study of its velocity field is required.\\

{\bf NGC4826}  is an Sab-type galaxy with a thick dust lane northeast of the nucleus. The
rotation of the outer gas disk in NGC 4826 is opposite to that of the inner one (Braun et
al. 1994). A detailed study of the $H_{\alpha}$ kinematics shows that the transition
between the two disks occurs near the dust lane at a distance $50''<r<70''$ from the
center (Rubin 1994). The  velocity $v_{max~disk}$ calculated from the R-band photometric
parameters of the disk (see Table \ref{table3}) and the $(B-V)$ and $(B-R)$ colors
beyond the dust lane  turns out to be higher than the  maximum rotational
velocity $v_{max}$ (see Table \ref{table4}). This conclusion remains valid even
if we use the model by Bell and de Jong (2001) with lower metallicity ($Z =
0.008$). Thus, our photometric data agree with the abnormally
low $M/L$ ratio in this galaxy.\\

{\bf NGC 5347} is an Sab-type galaxy with a bar. The
$PV-$ diagram constructed from optical observations
has a large scatter of points (Marquez et al. 2004)
and, therefore, does not allow $v_{max}$ to be determined
reliably. However, the galaxy's rotational velocity
estimated from the H\,{\small I} line width ($W_{20}$  from the RC3
catalog by de Vaucouleurs et al. 1991) turns out to be
lower than that for the disk component of the rotation
curve determined from the radial $R-$band surface
brightness profile (after the subtraction of the bulge
contribution) using the color indices of the galaxy's
outer regions (see Table \ref{table4}). Thus, an abnormally
low mass-to-light ratio for the disk can actually take
place in this case. This object requires a careful
study and, first of all, it is necessary to obtain a more
extended rotation curve.\\

{\bf NGC 5921} is an Sbc-type galaxy with a noticeable bar surrounded by a ring. As for
NGC 4214, the inclination estimates for this galaxy are contradictory. According to
Hernandez et al. (2005), the isophotes ellipticity leads to $i=36.5^o$. Our
photometric data give $i=43^o$ (see Fig. 2), which agrees with the inclination determined
kinematically from $H_{\alpha}$ observations (Hernandez et al. 2005). In Table
\ref{table4}, the rotational velocity $v_{max~disk}$ determined photometrically in the
same way as for NGC 5347 (see above) is compared with the two values of rotation velocity
$v_{max}$ found from $H_{\alpha}$ measurements for $i=43^o$ and $i=36.5^o$
for radial distance corresponding to $v_{max~disk}$. As we see from Table
\ref{table4}, for $i=36.5^o$ $v_{max~disk}$ is lower than $v_{max}$, while the
opposite conclusion is true for $i=43^o$. Since our inclination estimates agree with the
latter value, an abnormally low mass-to-light ratio for the disk of NGC 5921
remains quite possible. However, an underestimation of the circular rotational
velocity caused by the bar cannot be ruled out. This is suggested by the fact
that the rotational velocity estimated from the H\,{\small I} line width (from the RC3
catalog) for $i=43^o$ is $v_{HI}=136$ $km s^{-1}$, which exceeds
$v_{max~disk}$ (see Table \ref{table4}). Thus, the abnormally low $M/L$ for NGC 5921 is
most likely related to noncircular motions and not to stellar composition anomalies.\\
{\bf NGC 6814} is an SABb-type galaxy with a ring. The inclination
of NGC 6814 is low (the photometric axial ratio gives $i=17^o$).
However, $i=8^o$ is required for $v_{max}$ determined from the
rotation curve in the H\,{\small I} line to become approximately equal to
$v_{max~disk}$. Such a low inclination does not correspond to the
photometric estimates (see Fig. 2). The presence of a bar
hardly may change the result, because  its presence is
evident only in the central region ($R < 1$ kpc), while the disk
component of the rotation curve has a maximum at $R \approx 3$ kpc.
Therefore, the conclusion about an abnormally low $M/L$ of this galaxy remains preferable.\\

{\bf NGC 7743} is an S0-a galaxy with a bar. As in the previous two cases, the abnormally
low $M/L$ for the disk may be caused by the influence of the bar on the rotation velocity. It may also partially be attributed to the galaxy's low
metallicity ($Z = 0.008$ according to Katkov et al. 2011). In Table \ref{table4},
the two values of $v_{max~disk}$ obtained photometrically for two different metallicities
 are compared with $v_{max}$ reached at $R = 5$ kpc. For $Z = 0.008$, the velocities
$v_{max~disk}$ and $v_{max}$ turn out to be close (see Table \ref{table4}). Consequently,
as has been noted above, the abnormally low $M/L$ in NGC 7743 is not
evident if to take into account a low metallicity of the stellar population and
the influence of the bar on the
$v_{max}$ estimate. \\

{\bf UGC 03685} is an Sb-type galaxy. Just as
NGC 6814 and NGC 5921, it has a bar and a ring,
while having a small inclination. The inclination
determined from the $H_{\alpha}$  kinematics is very uncertain
($i=12^o\pm 16$) (Epinat et al. 2008). According to our
photometric data, $i=40^o\pm 15$  (see Fig. 2), which is
consistent with the NED ($33^o$) and Hyperleda ($55^o$)
estimates. In Table \ref{table4}, the disk rotational velocity
$v_{max~disk}$ estimated photometrically is compared with
the total rotational velocities $v_{max}$ corresponding to
$i_1=12^o$ and $i_2=40^o$. The conclusion about an
abnormally low mass-to-light ratio for the disk of
UGC 03685 is confirmed if we take the photometrically
estimated inclination $i_2$ and is not confirmed
for $i_1$. If we take into account the large error in the
kinematic inclination estimate, then the conclusion
about an abnormally low $M/L$ remains possible,
though uncertain.\\

Table \ref{table4} contains the following data:\\
(1) galaxy name;\\
(2) maximum rotational velocity of the disk-related component
of the rotation curve $v_{max~disk}$ determined photometrically based on the Bell, de Jong (2001) models;\\
(3) maximum rotational velocity $v_{max}$ derived from direct observations (for
NGC 7743, NGC 5921, NGC 6814, and UGC 3685, this is the velocity of rotation at
the distance $R=2r_d$, where
the disk contribution to the rotation curve is maximal);\\
(4) reference to the source of the rotation curve;\\
(5) note;\\
(6) photometrically determined mass-to-light ratio
for the stellar population of the disk (in the R- band)
from the model by Bell, de Jong (2001);\\
(7) total dynamical mass-to-light ratio for the galaxy in the R-band within the optical
radius $R = 4r_d$ (for NGC 1569 and NGC 4214 it is within $R_{25}$); for NGC
1569 the second row of the table gives the total $M/L_{Ks}$ ratio;\\
(8) total mass-to-light ratio for the galaxy $M/L_R$ calculated from the models by Bell,
de Jong (2001) and the total  $(B- R)_0$ color (for NGC 1569, the second row gives
$M/L_{Ks}$ calculated from the
total $(J- Ks)_0$ color).\\
\begin{table}[h!]
\caption{Comparison of $v_{max~disk}$ with $v_{max}$ and $M/L$ estimates \label{table4}}
%\begin{center}
   \begin{tabular}{|c|p{1.6cm}|p{1.4cm}|c|c|c|c|p{1.7cm}|}
  \hline
{\small Galaxy}&$v_{max~disk}$, {\small km s}$^{-1}$ & $v_{max}$, {\small km} $s^{-1}$  & Ref. & Note&$(M/L_R)_d$&$M/L_{R~dyn}$&$M/L_{R~mod}$ \\
 (1) & (2)&(3)&(4)&(5)&(6)&(7)&(8)\\
\hline
{\small NGC1569}&60&43&\cite{stil}&R&0.16&0.146&0.26\\
{\small NGC1569}&45&43&\cite{stil}&Ks&0.49&0.8&0.49\\
\hline
{\small NGC4016}&73&78&\cite{ngc4016}&---&0.8&1.53&0.86\\
\hline
{\small NGC4214}&75&42&\cite{ngc4214}&$i=40^o$&0.8&0.3&0.73\\
{\small NGC4214}&75&79&\cite{ngc4214}&$i=20^o$&0.8&1.06&0.73\\
\hline
{\small NGC4826}&256&154&\cite{ngc4826}&$Z=0.02$&2.25&1.06&2.24\\
{\small NGC4826}&241&154&\cite{ngc4826}&$Z=0.008$&2&1.06&2.02\\
\hline
{\small NGC5347}&111&56&\cite{rc3},\cite{Marquez2004}&$Z=0.02$&1.5&0.65&2.00\\
{\small NGC5347}&103&56&\cite{rc3},\cite{Marquez2004}&$Z=0.008$&1.3&0.65&1.79\\
\hline
{\small NGC5921}&139&111&\cite{ngc5921}&$i=43^o$&1.5&1.75&1.77\\
{\small NGC5921}&139&127&\cite{ngc5921}&$i=36.5^o$&1.5&2.3&1.77\\
\hline
{\small NGC6814}&148&120&\cite{ngc6814}&---&1.75&1.07&2.17\\
\hline
{\small NGC7743}&135&118&\cite{katkov}&$Z=0.02$&2.28&2.3&2.37\\
{\small NGC7743}&125&118&\cite{katkov}&$Z=0.008$&2&2.3&2.15\\
\hline
{\small UGC03685}&86&101&\cite{ugc3685},\cite{Ghasp}&$i=12^o$&1.25&3.37&1.69\\
{\small UGC03685}&86&33&\cite{ugc3685},\cite{Ghasp}&$i=40^o$&1.25&0.35&1.69\\
\hline
\end{tabular}
\end{table}

\section{CONCLUSIONS}
We present the results of our surface photometry in the B, V and R bands for
nine disk galaxies in which the discrepancy between low M/L ratio  and
color of stellar population is suspected. We obtain the photometric profiles, the
radial profiles of the color indices, the position angle, and the flattening of isophotes. 
We decomposed the images using the estimation of photometric parameters for
individual components (bulge, bar and disk) of the galaxies.  We apply two methods
for the dynamical and photometric disk mass estimates: by checking whether the
colors agree with the total $M/L$ ratios of the galaxies, which may be considered as an
upper limit of $M/L$ for the stellar population, and by comparing the observed rotational
velocities of the galaxies with the maximum circular velocity of the disk expected
from its
photometric parameters. Our results show that there are no obvious
contradictions between the dynamical and photometric mass estimates in most
considered objects because the
abnormally low dynamical $M/L$ estimates can be naturally explained by the uncertainty in
determining the rotational velocity or by the errors in photometry. At the
same time our
photometric data for NGC 4826, NGC 6814, NGC 5347, and with lesser confidence for UGC
03685 and NGC 4214 suggest that these galaxies may actually be too ''light'' for their
luminosity, and hence may have the IMF defficient by
low-mass stars. The presence of dark matter in the galaxies which was
ignored in our study would only increase the discrepancy between the dynamical and photometric
$M/L$ estimates. The galaxies we discuss require a more complete study: first of all, more
careful measurements of the rotational velocity far from the center are
needed.

The low values of the mass-to-light ratio in galaxies, even if this is not a
result of anomalous stellar population, suggest the low
dark-to-luminous matter mass fraction in them.\\

\section{ACKNOWLEDGMENTS}
We wish to thank R. Swaters who kindly provided
the brightness profile of NGC 1569. We also wish
to thank R.E. de Souza and D.A. Gadotti for the
opportunity to use the BUDDA code. We are grateful
to Hyperleda support team for the opportunity to
use this database.
Based on observations obtained with the Apache Point Observatory 0.5-meter
telescope, which is owned and operated by the Astrophysical Research
Consortium.\\
This work was supported by Russian Foundation for Basic Research, grant 11-02-12247.

\end{document}